\documentclass[sts]{imsart}

\input{packages_and_macros.tex}

\RequirePackage[round,authoryear]{natbib}

\begin{document}

\begin{frontmatter}

\title{Exponential-Family Models of Random Graphs: Inference in Finite-, Super-, and Infinite-\\Population Scenarios}

\runtitle{ERGMs}

\begin{aug}

\author{\fnms{Michael} \snm{Schweinberger}},
\author{\fnms{Pavel N.} \snm{Krivitsky}},
\author{\fnms{Carter T.} \snm{Butts}},\\
\hspace{-.12cm}
\and
\author{\fnms{Jonathan} \snm{Stewart}}

\address{Michael Schweinberger is Assistant Professor, Department of Statistics, Rice University, Houston, Texas 77005, U.S.A.\ (email: m.s@rice.edu). 
Pavel N.\ Krivitsky is Lecturer, Department of Statistics, University of New South Wales, Sydney, New South Wales 2052, Australia (email: p.krivitsky@unsw.edu.au).
Carter T.\ Butts is Professor, Departments of Sociology, Statistics, Computer Science, and Electrical Engineering and Computer Science, University of California, Irvine, California 92697, U.S.A.\ (email: buttsc@uci.edu).
Jonathan Stewart is graduate student, Department of Statistics, Rice University, Houston, Texas 77005, U.S.A.\ (email: jrs6@rice.edu).}

\thankstext{t1}{We are indebted to Catalina Obando and Fabrizio {D}e {V}ico {F}allani for sharing the human brain network data used in Section \ref{application}.
In addition, 
we are grateful to Johan Koskinen and the Department of Statistics at the University of Stockholm, Sweden for making the seminal works on pseudo-likelihood-based versus likelihood-based inference from the 1990s accessible.
Last, 
but not least, 
we would like to thank an associate editor and three anonymous referees for helpful suggestions, 
and Noel Cressie, Mathias Drton, Mark S.~Handcock, Eric D.~Kolaczyk, Martina Morris, and Thomas Richardson for insightful discussions.
Michael Schweinberger and Jonathan Stewart were partially supported by NSF awards DMS-1513644 and DMS-1812119.
Carter T.~Butts was partially supported by ARO award W911NF-14-1-0552 and NSF award DMS-136142.} 

\runauthor{Schweinberger et al.}

\end{aug}

\begin{abstract}
Exponential-family Random Graph Models (ERGMs) constitute a broad statistical framework for modeling both sparse and dense random graphs with features such as 
short- or long-tailed degree distributions,
covariate effects,
and a wide range of complex dependencies.
Special cases of ERGMs include
network equivalents of generalized linear models (GLMs),
Bernoulli random graphs,
$\beta$-models, 
$p_1$-models, 
and models related to the Markov random fields employed in spatial statistics and image analysis.
While ERGMs are widely used in practice,
questions have been raised about their theoretical properties.
These include concerns that some ERGMs are near-degenerate and that many ERGMs are non-projective.
To address such questions,
careful attention must be paid to model specifications and their underlying assumptions,
and to the inferential settings in which models are employed.
As we discuss, 
near-degeneracy can affect simplistic ERGMs lacking structure, 
but well-posed ERGMs with additional structure can be well-behaved.  
Likewise, 
lack of projectivity can affect non-likelihood-based inference,
but likelihood-based inference does not require projectivity.
Here, 
we review well-posed ERGMs along with likelihood-based inference for typical ERGM settings.
We first clarify the core statistical notions of ``sample'' and ``population'' in the ERGM framework,
separating the process that generates the population graph from the observation process.
We then review likelihood-based inference in finite-,
super-,
and infinite-population scenarios.
We conclude with consistency results, and an illustrative application to human brain networks.

{\em Keywords:} social network, exponential-family random graph model, ERGM, model degeneracy, projectivity.
\end{abstract}

\end{frontmatter}

\defcitealias{ShRi11}{\citet{ShRi11}}

\tableofcontents

\section{Introduction}
\label{sec:introduction}

The statistical analysis of network data is an emerging area of statistics with applications in epidemiology (e.g., disease-transmission networks),
neuroscience (e.g., human brain networks),
political science (e.g., insurgencies, terrorist networks),
economics (e.g., financial markets),
sociology (e.g., social networks),
and computer science (e.g., Facebook, Twitter) \citep{Ko09a}.
Core to the statistical analysis of network data is the development of random graph models.
There is a large and growing body of literature on random graph models \citep{goldenberg2009survey,Fien12,HuKrSc12,latentspace},
including exchangeable random graph models \citep[e.g.,][]{DiJa08,BiCh09,CaCaBr16,CaFo17,Ja17,CrDe16,CrDe15,crane2018probabilistic},
stochastic block models \citep[e.g.,][]{NkSt01,BiCh09,RoChYu11},
latent space models \citep[e.g.,][]{HpRaHm01,HaRaTa07,sewell2015latent},
and many other random graph models \citep[e.g.,][]{Ho05,RaFrRa15,Ho18,Foetal18}.

Most such models seek to address the twin challenges of modeling complex and dependent network data:
the presence of \emph{heterogeneity} (different subgraphs may have different properties) and \emph{dependence} (the presence or absence of an edge may be dependent upon the presence or absence of other edges in the graph).  
The latter,
dependence,
has proven to be the more challenging of the two.
Indeed,
network data can exhibit a wide range of complex dependencies, 
the most famous of which is transitivity or triadic closure bias \citep[e.g.,][]{WsFk94},
which has been found in human brain networks \citep[e.g.,][]{SiMoLa12,brain16,brain17}, 
social networks \citep{HpLs70,HpLs72,HpLs76},
and other network data.  
Other examples include degree heterogeneity \citep{Hu08} and suppression of short chordless cycles \citep{bearman.et.al:ajs:2004}.
We focus here on models designed to capture a wide range of complex dependencies,
including---but not limited to---local effects such as triadic closure.

A broad statistical framework that has turned out to be useful for testing and modeling complex dependencies in areas as diverse as physics, 
neuroscience,
artificial intelligence,
spatial statistics, 
and other areas of statistics are exponential-family models for dependent random variables \citep{BN78,Br86}.
Such models have long been used in statistics and related areas,
classic examples being Ising models in physics \citep{Ising} and discrete Markov random fields in spatial statistics \citep{Bj74,Cr93,St99} and machine learning \citep[e.g.,][]{RaWaLa10,Yaetal15}.
In the network science literature,
discrete exponential-family models for dependent random variables are known as Exponential-Family Random Graph Models (ERGMs) \citep{FoSd86,WsPp96,ergm.book,Ha13}.
These models are versatile and,
when properly specified,
are capable of modeling both sparse and dense random graphs with features such as
short- or long-tailed degree distributions,
covariate effects,
and a wide range of complex dependencies \citep[e.g.,][]{ergm.book,Ha13}.
Some notable special cases of ERGMs include:\s
\bi
\item ``network'' GLMs \citep{MpNj83}, 
including network logistic regression and other forms of network regression \citep{krackhardt:sn:1988};
\item Bernoulli random graphs \citep{Gi59,ErRe59,ErRe60};
\item categorical data models \citep[e.g., U$|$MAN models,][]{WsFk94};
\item $\beta$-models \citep{DiChSl11};
\item $p_1$ models \citep{HpLs81};
\item canonical exponential-family models with Markov dependence \citep{FoSd86},
which are related to Ising models in physics \citep{Ising},
Markov logic networks in artificial intelligence \citep{RiDo06},
Markov random fields in spatial statistics \citep{Bj74,Cr93,St99},
and undirected graphical models \citep[e.g.,][]{La96,RaWaLa10};
\item and curved exponential-family models \citep{SnPaRoHa04,HuHa04,HuGoHa08}.\s
\ei

\alert{The advantages of exponential-family representations of random graph models are many.
Some of the more important ones are:\s
\begin{enumerate}
\item \emph{Language.} Exponential families provide a convenient language for formulating ideas about network data and dependencies therein. 
In so doing,
the language of exponential families facilitates the construction of models for complex and dependent network data.
\item \emph{Unifying statistical framework.} The exponential-family framework is a unifying statistical framework that includes a wide range of random graph models,
as mentioned above.
The unifying exponential-family framework has computational, theoretical, and practical advantages.
\item \emph{Computational advantages.} Exponential families have useful convexity properties,
e.g.,
the natural parameter space of exponential families is a convex set and the negative loglikelihood function is a strictly convex function on the natural parameter space \citep{Br86}.
These convexity properties imply that maximum likelihood estimation of natural parameters is a convex minimization program with a unique solution,
provided that there exists a solution \citep{Ha03p,fienberg-2008}.
In addition,
exponential families admit data reduction by sufficient statistics,
and exponential-family likelihood functions depend on the data through minimal sufficient statistics \citep{Fi34}.
As a consequence,
likelihood-based estimation algorithms are agnostic to the structure of the sample space of the network data---which can be a large and discrete set---as long as observed and expected minimal sufficient statistics can be computed exactly or approximately \citep{KrBu16}.
Thus, 
the unifying statistical framework helps design statistical algorithms that can estimate a wide range of random graph models with exponential-family representations,
and helps implement them in user-friendly statistical software.
In fact,
there are 20 ERGM-related \texttt{R} packages,\footnote{\label{ergm.r}The 20 ERGM-related \texttt{R} packages found at \texttt{CRAN.R-project.org}: \texttt{Bergm}, \texttt{blkergm}, \texttt{btergm}, \texttt{dnr}, \texttt{EpiModel}, \texttt{ergm}, \texttt{ergm.count}, \texttt{ergm.ego}, \texttt{ergm.graphlets}, \texttt{ergm.rank}, \texttt{ergm.userterms}, \texttt{ergmharris}, \texttt{fergm}, \texttt{GERGM}, \texttt{gwdegree}, \texttt{hergm}, \texttt{mlergm}, \texttt{statnetWeb}, \texttt{tergm}, and \texttt{xergm}.}
with tens of thousands of downloads a month among them.
It is worth noting that likelihood-based methods require approximations of the likelihood function when it is intractable (a problem shared with discrete Markov random fields in spatial statistics and machine learning),
but there are multiple general-purpose methods for approximate likelihood-based inference (see Section \ref{computational.methods}).
\item \emph{Theoretical advantages.} Among the theoretical advantages is the fact that the exponential-family framework helps construct statistical theory for a wide range of random graph models that have exponential-family representations.
The seminal work of \citet{Be72} and \citet{Po88} and others demonstrates that statistical theory is easier for exponential families than non-exponential families,
and statistical theory for dependent network data is likewise facilitated by exponential families \citep[as demonstrated in Section 3.1 of][]{ScSt16}.
Network data are complex enough,
and it therefore makes sense to keep statistical theory as simple as possible.
The exponential-family framework enables theoreticians to do so.
\item \emph{Practical advantages.} Among the practical advantages is the fact that ERGMs include GLMs for independent network data---e.g.,
logistic regression models---as special cases and may be viewed as GLMs for independent as well as dependent network data.
Many network scientists are familiar with logistic regression models,
which facilitates the interpretation of results, in that estimates of parameters may be interpreted in terms of conditional log odds and log odds ratios \citep[e.g.,][]{Hu08,ergm.book,StScBoMo18}.\s
\end{enumerate}
}
In addition,
exponential-family models can be used as building blocks to construct more complex models,
e.g.,
by using them as building blocks in stochastic block models to capture transitivity and other complex dependencies within blocks \citep{ScHa13}.

In contrast to latent space models \citep[e.g.,][]{HpRaHm01} and other latent variable models,
ERGMs directly specify dependence by incorporating network features as sufficient statistics, 
providing a direct link between the structural properties on which network science theories are based \citep{WsFk94} and the resulting models.
By contrast, 
most latent variable models represent dependence among edges as the result of underlying heterogeneity in edge probabilities (represented by latent structure), 
in some cases making interpretation less direct.  
While both approaches have useful applications \citep[and can be combined, see, e.g.,][]{Ko09c,ToMu14,thiemichen2016bayesian},
the ability to directly and parsimoniously model dependencies of scientific interest helps explain the popularity of Ising models in physics and ERGMs in network science.
Indeed,
well-posed ERGMs have found widespread application,
ranging from the study of the human brain \citep[e.g.,][]{SiMoLa12,brain16,brain17} and epidemics \citep[e.g.,][]{GrWeHu11} to differential privacy \citep[e.g.,][]{karwa2017sharing}, transient structure in intrinsically disordered proteins \citep{grazioli.et.al:f:2019}, and networks of radical environmentalists \citep{AlBa15}.

While ERGMs are widely used in practice,
questions have been raised regarding their theoretical properties.
To a large extent, 
these questions reflect a lack of clarity about the construction of ERGMs and the underlying assumptions of particular model specifications,
and about the inferential settings in which ERGMs are employed.
ERGMs may be used, 
among other things, 
to draw conclusions regarding the state of a single graph based on sampled subgraphs,
the nature of a generative process that produces graphs of fixed size, 
or the properties of a generative process that produces sequences of graphs of arbitrary size.  
These scenarios are distinct, 
and have implications for both model specification and inference.  
The main goal of our paper is to clarify these distinct uses, 
with an eye to facilitating more principled inference.  
We argue that concerns raised regarding the uses of ERGMs can be addressed by paying close attention to model specifications and their underlying assumptions along with the inferential settings in which models are used, discussing likelihood-based inference for well-posed ERGMs in finite-, super-, and infinite-population scenarios. 
We review consistency results for likelihood-based estimators,
and demonstrate likelihood-based inference in super-population scenarios by an illustrative application to human brain networks.

\subsection{Outline}

To prepare the ground for the remainder of our paper,
we first describe and address two questions that have been raised about the theoretical properties of ERGMs.
An examination of these questions demonstrates the need for proper statistical language to clarify the core statistical notions of ``sample'' and ``population'' in the ERGM framework, and shows that a failure to use proper statistical language may result in misleading statistical conclusions.
In the remainder of the paper,
we provide a review of ERGMs that\s
\begin{itemize}
\item introduces exponential-family parameterizations of random graph models (Section \ref{ergm});
\item describes and addresses two questions that have been raised about the theoretical properties of ERGMs (Section \ref{concern});
\item reviews ERGMs with additional structure, 
which helps construct ERGMs with attractive theoretical properties (Section \ref{ergms.additional.structure});
\item clarifies the core statistical notions of ``sample'' and ``population'' in the ERGM framework,
separating the complete-data generating process (the process that generates the population graph) from the incomplete-data generating process (the observation process) (Section \ref{generating.processes});
\item distinguishes statistical inference for graphs of fixed size,
including finite- and super-population inference,
and statistical inference for sequences of graphs of increasing size (Section \ref{generating.processes});
\item describes likelihood-based inference given incomplete data,
generated by ignorable incomplete-data processes (Section \ref{sec:likelihood});
\item discusses consistency and asymptotic normality of likelihood-based estimators in finite-, super-, and infinite-population scenarios (Section \ref{ergm.consistency});
\item demonstrates likelihood-based inference for well-posed ERGMs by an application to human brain networks (Section \ref{application}).
\end{itemize}

\s

\subsection{Topics not covered}
\label{topics}

The study of ERGMs is a rich area of statistical network analysis, 
with many substantive, computational, and statistical topics that could be addressed.  
As our focus is the specification of well-posed ERGMs and accompanying inference scenarios, 
there are many other issues that space does not allow us to address.  
Here, 
we mention some of these topics, 
along with pointers to related work.

\subsubsection{Computational methods for ERGMs}
\label{computational.methods}

We do not cover computational methods for ERGMs.  
While computational methods are important in their own right, 
they are separable from the theoretical issues discussed here, 
and discussions can be found elsewhere in the literature.  
Many computational methods for ERGMs are closely related to computational methods for other discrete exponential-family models for dependent random variables,
such as Ising models in physics and discrete Markov random fields in spatial statistics.
While they differ in their particulars, such models pose similar computational challenges---particularly, likelihood functions with intractable normalizing constants and posteriors with doubly intractable normalizing constants---and therefore similar solutions are applicable to many of them. 
Without providing additional details,
we mention here the main classes of computational approaches that have been explored to date:\s
\begin{itemize}
\item Pseudo-likelihood \citep{StIk90,LeCrDe17} and composite-likelihood methods \citep{AsLiIhSm10,Kr17},
contrastive divergence \citep{AsLiIhSm10},
and approximations based on graph limits \citep{HeZh15};
note that methods which are not based on the likelihood function need to be used with caution,
as explained in Section \ref{clarification.likelihood}.
\item Stochastic approximation methods for maximum likelihood estimation \citep{Sn02,SnDu02,Jin2013p927}.
\item Monte Carlo maximization methods for maximum likelihood estimation \citep{GeTh92,Ha03p,HuHa04,HuHuHa12,OkGe12,YaRiFi14,Kr17,Byetal18}.
\item Bayesian Markov chain Monte Carlo methods \citep{Ko04,moller-etal-2006a,murray2006,CaFr09,Ev12,AtLaRo12,CaFr12,JiYuLi13,Liang_NC2013,WaAt13,CaMi2015,Lyne15,LiJi16,PaHa18}.
\item Incomplete-data maximum likelihood and Bayesian estimation of ERGMs based on sampled data and missing data \citep{GiHa06,HaGi09,KoRoPa09,Sn10,PaRoSn13,KrMo17,karwa2017sharing}.
\item To generate draws of graphs from ERGMs,
Markov chain Monte Carlo \citep{Sn02,ergmtermsjss} is the workhorse approach,
though perfect sampling \citep{Bu12} and non-Markov chain Monte Carlo approximate sampling approaches \citep{butts:jms:2015} have also been developed.
\end{itemize}

\subsubsection{Mixtures of ERGMs}
\label{mixture.ergms}

As our focus is on ERGMs per se, 
we do not cover random graph models that use ERGMs as building blocks,
such as mixtures of ERGMs.
Most mixtures of ERGMs are finite or infinite mixtures of Bernoulli random graphs.
Examples are:\s
\begin{itemize}
\item Stochastic block models \citep[e.g.,][]{NkSt01,ABFX08,BiCh09,BiChLe11,ChWoAi12,RoChYu11,zhao2012,AmChBiLe13,LeRi13,Jin15,Gao15,ZhZh16,BiVoRo17,SeCh18}.
\item Latent space models \citep[e.g.,][]{HpRaHm01,ScSn03,HaRaTa07,RaNiHoYe12,STMu13,TaSuPr13,sewell2015latent}.
\item Other latent variable models \citep[e.g.,][]{Du95,DuSnZi04,GiSw04,Ho05,Ho08,Ho09,Ho18,FoHo15,Foetal18}.
\end{itemize}
In many applications of such models,
interest centers on the latent variable structure,
e.g.,
the unobserved block structure in applications of stochastic block models.
Although we do not focus on such models, many of the issues we discuss---in particular the inference scenarios reviewed here---may inform treatment of ERGM mixtures as well.
 
\subsubsection{Exchangeable random graph models and other models}
\label{exchangeable}

Various other models have been proposed,
such as exchangeable random graphs---random graphs invariant to the labeling of nodes \citep{DiJa08,BiCh09} or the labeling of edges \citep{CaCaBr16,Ja17,CrDe16,CrDe15,crane2018probabilistic};
scale-free networks \citep[e.g.,][]{BaAl99,Boetal01,scalefree};
and other models \citep[e.g.,][]{rapoport:sn:1979}.

The literature on such models is large and diverse, as are their motivations and properties, and their relationships to ERGMs have not been studied in depth.
We hence do not attempt to treat these models here, 
but refer readers to the above-cited papers and other related literature  \citep[e.g.,][]{DiJa08,Lo12,OrRo15,VeRo15,VeRo16,CaFo17,Boetal17,Ja17,LaRiSa17}.
In the case of exchangeable random graphs, we do discuss one notable example:
the edge-exchangeable models of \citet{CrDe16}.
Edge-exchangeable models assign equal probabilities to all edge-labeled graphs which are isomorphic up to relabeling \citep[Definition 3.1,][]{CrDe16}.
A special case of the generic modeling framework was elaborated by \citet{CrDe16},
the ``Hollywood model.''
The motivating example of the Hollywood model concerns actors linked by movies,
although the model is not restricted to the motivating example.
While such models hold promise,
it is an open question whether, 
and how, 
those models can be used to address one of the key challenges of statistical network analysis:
network data are dependent data,
and testing and modeling dependencies in network data is of great interest.
The range of dependencies captured by those models has not been characterized to date,
and more research is needed to determine how such models can be used to test and model complex dependencies in network data,
such as transitivity.  

\section{Exponential-family parameterizations of random graph models}
\label{ergm}

\alert{To introduce exponential-family parameterizations of random graph models,
which are known as Exponential-family Random Graph Models (ERGMs),
we consider a finite population of nodes $\verts = \{1, \dots, N\}$ ($N \geq 2$).
The nodes $i \in \verts$ may have attributes (e.g., age).
We denote the collection of attributes of population members by $\attrs_{\verts} \in \attrf_{\verts} \subseteq \mR^q$.
In addition to attributes,
the nodes $i \in \verts$ may be connected to other nodes $j \in \verts$ by edges.
In the simplest case,
edges are undirected and self-edges are excluded,
although many of the key ideas we discuss can be extended to directed edges.
Edges between nodes $i\in\verts$ and $j\in\verts$ are considered random variables and can take on weights,
denoted by $\Edge{i,j}$.
The weights of edges $\Edge{i,j}$ can take on values in the set\s
\bi
\item $\netset = \{0, 1\}$, 
where $0$ indicates the absence of an edge whereas $1$ indicates the presence of an edge;
\item $\netset = \{0, 1, \dots\}$, 
where $0, 1, \dots$ are counts of the number of interactions or other relational events;
\item $\netset = \mR$, 
indicating financial transactions or other relational events with real-valued outcomes.\s
\ei
The vast majority of ERGM-related publications focuses on random graphs $\Net = (\Edge{i,j})_{i\in\verts < j\in\verts}$ with sample spaces of the form $\netset_{\verts} = \{0, 1\}^{\binom{|\verts|}{2}}$,
but there are exponential-family models of random graphs with sample spaces of the form 
$\netset_{\verts} = \{-1, 0, 1\}^{\binom{|\verts|}{2}}$ \citep{HuDu12u},
$\netset_{\verts} = \{0, 1, \dots\}^{\binom{|\verts|}{2}}$ \citep{Kr11},
and $\netset_{\verts} = \mR^{\binom{|\verts|}{2}}$ \citep{DeCr12},
as well as random graphs where edges are ranks \citep{KrBu16}, 
categorical with unordered categories \citep{RgPpWs99} or ordered categories \citep{CaGo20m}, 
and multivariate combinations thereof \citep{PpWs99,LaPa99m,KrMa19e}.
In fact, 
there are entire {\tt R} packages devoted to ERGMs for random graphs with sample spaces of the form $\netset_{\verts} = \{0, 1, \dots\}^{\binom{|\verts|}{2}}$ and $\netset_{\verts} = \mR^{\binom{|\verts|}{2}}$.
}

\alert{To cover random graphs with sample spaces of the form $\netset_{\verts} = \{0, 1\}^{\binom{|\verts|}{2}}$,
$\netset_{\verts} = \{0, 1, \dots\}^{\binom{|\verts|}{2}}$,
$\netset_{\verts} = \mR^{\binom{|\verts|}{2}}$,
and other sample spaces that have been explored in the literature,
we consider exponential families of densities with respect to a $\sigma$-finite reference measure $\mref$ with support $\netset_{\verts}$,
specified by a sufficient statistic $\genstats: \attrf_{\verts} \times \netset_{\verts} \mapsto \mR^{\nnatpar}$ and a map $\cnmap: \paramf \times \verts \mapsto \mR^{\nnatpar}$ with $\paramf\subseteq\en\{\}{\params\in\mR^p: \lnormc(\params, \verts)<\infty}$:
\beq
\displaystyle\frac{\wrt \M_{\verts,\cnmap(\params, \verts)}}{\wrt\mref}(\net_{\verts})
\eqs \exp\en(){\langle\cnmap(\params, \verts),\, \genstats(\attrs_{\verts},\, \net_{\verts})\rangle-\lnormc\en(){\params, \verts}},
\;\; \net_{\verts}\in\netset_{\verts},\label{eq:ergm}
\eeq
where $\langle\cnmap(\params, \verts),\, \genstats(\attrs_{\verts},\, \net_{\verts})\rangle$ denotes the inner product of natural parameter $\cnmap(\params, \verts)$ and sufficient statistic $\genstats(\attrs_{\verts},\, \net_{\verts})$ and
\beq
\lnormc\en(){\params, \verts}
\eqs \log\dint_{\netset_{\verts}} \exp\en(){\langle\cnmap(\params, \verts),\, \genstats(\attrs_{\verts},\, \net_{\verts}^\prime)\rangle}\wrt\mref\en(){\net_{\verts}^\prime}.
\eeq
To present the key ideas in the simplest possible setting,
we focus henceforth on the simplest and most common case:
exponential-family models of random graphs with sample spaces of the form $\netset_{\verts} = \{0, 1\}^{\binom{|\verts|}{2}}$,
in which case 
\beq
\M_{\verts,\cnmap(\params, \verts)}(\net_{\verts})
\eqs \exp\en(){\langle\cnmap(\params, \verts),\, \genstats(\attrs_{\verts},\, \net_{\verts})\rangle-\lnormc\en(){\params, \verts}}\, \mref(\net_{\verts}),
\;\; \net_{\verts}\in\{0, 1\}^{\binom{|\verts|}{2}},
\eeq
where
\beq
\lnormc\en(){\params, \verts}
\eqs \log\dsum_{\net_{\verts}^\prime \in\, \netset_{\verts}} \exp\en(){\langle\cnmap(\params, \verts),\, \genstats(\attrs_{\verts},\, \net_{\verts}^\prime)\rangle}\, \mref\en(){\net_{\verts}^\prime}.
\eeq
All of these quantities can depend on the population of nodes $\verts$.
}

\alert{The generic exponential-family framework may be intimidating and the reader may question the high level of abstraction and generality of the above definition,
not the least the fact that all quantities are allowed to depend on the population of nodes $\verts$.
A simple example may help demonstrate why all quantities can depend on $\verts$,
and why it is desirable to allow them to depend on $\verts$.
Consider the family of sparse Bernoulli$(\pi_{|\verts|})$ random graphs with size-dependent edge probabilities $\pi_{|\verts|} = \logit^{-1}(\theta - \log |\verts|)$ ($\param\in\mR$),
with probability mass function
\beq
\M_{\verts,\eta(\param,\verts)}(\Net_{\verts} = \net_{\verts})
\;=\; \exp\left(\eta(\param, \verts) \displaystyle\sum_{i\in\verts\, <\, j\in\verts} \edg{i,j} - \psi(\param, \verts)\right)\, \mref(\net_{\verts}),
\eeq
where the support of the reference measure depends on $\verts$:
\beq
\mref(\net_{\verts}) \;=\;
\begin{cases}
1 & \mbox{ if } \net_{\verts} \in \{0, 1\}^{\binom{|\verts|}{2}}\s
\\
0 & \mbox{ otherwise,}
\end{cases}
\eeq
as do the natural parameter $\eta(\param, \verts) = \param - \log |\verts|$ and the log-normalizing constant $\psi(\param, \verts) = \log(1 + \exp(\param - \log |\verts|))$.
In other words,
all quantities depend on the population of nodes $\verts$.
Indeed,
the dependence on $\verts$ stems from the offset $\log |\verts|$,
which induces sparsity in Bernoulli random graphs.
We motivate sparsity and sparsity-inducing Bernoulli$(\pi_{|\verts|})$ random graphs in Section \ref{concern.nonprojective}.
Covariates can be included by using the logit link function $\logit(\pi_{|\verts|}) = \param - \log |\verts|$ and adding covariate terms,
as in logistic regression \citep{MpNj83}.
Covariate terms are reviewed in \citet{ergmtermsjss}.
Examples of other ERGMs with support $\netset_{\verts} = \{0, 1\}^{\binom{|\verts|}{2}}$ can be found throughout the paper.
ERGMs with other forms of support can be found in the literature cited above.
}

\section{Two questions raised about ERGMs and clarifications}
\label{concern}

Here, 
we describe two questions that have been raised regarding the theoretical properties of ERGMs, 
discuss the historical and mathematical context in which those questions arose,
and outline which lessons have been learned and how the associated issues have been addressed.  
The resulting discussion motivates a more careful look at the specification of ERGMs and inference for ERGMs, 
which are discussed in the following sections of our paper.

\subsection{Question I: Are non-trivial ERGMs near-degenerate?}
\label{concern.bad}

\subsubsection{Overview and history}

A common concern is that ERGMs with non-trivial dependence structure can be ill-behaved,
in the sense that ERGMs can be either near-degenerate or indistinguishable from Bernoulli random graphs with, 
in some cases, 
a phase transition between these two regimes \citep{St86,Jo99,HaJo99,Ha03p,BaBrSl08,fienberg-2008,BaBrSl11,Sc09b,Bu10,ChDi11,Me17,BhCh18w}.
Both near-degenerate ERGMs and near-Bernoulli models are problematic as models of network data:
near-degenerate ERGMs concentrate probability mass on small subsets of graphs,
such as near-complete graphs with almost all possible edges,
whereas near-Bernoulli random graphs induce vanishing dependence.
These properties render them useless in most applications,
although there are exceptions.
One exception is the behavior of some physical systems (e.g., networks of crystal contacts or amlyoid fibrils),
which resembles the behavior near-degenerate ERGMs---including the existence of phase transitions.
In such cases,
the behavior of near-degenerate ERGMs can be both realistic and desirable \citep{grazioli.et.al:jpcB:2019}.
However,
we here follow here convention and treat such behavior as undesirable for typical use cases.

The fact that some ERGMs are ill-behaved was first discovered by \citet{St86},
\citet{Jo99},
and \citet{HaJo99}.
We discuss the theoretical results of these pioneers in Section \ref{clarification.theoretical} along with more recent work.
In practice,
the undesirable properties of ill-behaved ERGMs went unnoticed at first,
in part because pseudo-likelihood-based methods---which masked the undesirable properties of ill-behaved ERGMs---were used to estimate them \citep{StIk90},
and in part because model assessment tools were unavailable.
The introduction of Markov chain Monte Carlo methods for generating draws of graphs from ERGMs in the 1990s and 2000s revealed that some models estimated by maximum pseudo-likelihood methods performed very poorly (generating, e.g., graphs with almost all possible edges when estimated from observed graphs with a moderate number of edges).
Originally,
this was thought to be due to poor estimation, 
and indeed maximum pseudo-likelihood estimators were shown to be inferior to maximum likelihood estimators \citep{DaDa93,DaDa99,CjDkDp98,CoDaDa02}.
Eventually,
Monte Carlo maximum likelihood estimators were developed \citep{Sn02,HuHa04},
along with simulation-based model assessment tools \citep{HuGoHa08}.
These developments revealed that, 
while some of the bad behavior did stem from inferior estimators, 
the bad behavior in other cases was inherent to the specified models,
as first pointed out by \citet{Sn02} and \citet{Ha03p} and anticipated by the work of \citet{St86}, \citet{Jo99}, and \citet{HaJo99}.
Since then,
statistical theory has shed more light on those undesirable properties, 
and has led to the development of improved model specifications.
We review the existing theoretical results in Section \ref{clarification.theoretical},
and improved model specifications in Section \ref{clarification.good}.

\subsubsection{Clarification: Most theoretical results are limited to simplistic ERGMs that lack structure}
\label{clarification.theoretical}

Since the 1980s \citep{St86,Jo99,HaJo99},
it has been known that some ERGMs are ill-behaved.
However,
two important points have been lost in more recent discussions of ill-behaved ERGMs.
First,
almost all theoretical results---discussed below---are limited to simplistic ERGMs which lack structure that could restrict interactions among edge variables and which have the same number of natural parameters,
regardless of how large $|\verts|$ is.
Such simplistic ERGMs resemble Ising models in physics without lattice structure or discrete Markov random fields in spatial statistics without spatial structure, 
and are of limited use in understanding large, complex and dependent networks.
Theoretical results based on such models are therefore limited in scope, and must be interpreted with careful attention to the underlying assumptions. 
As we shall discuss, 
most of those results do not generalize to ERGMs with additional structure.
Indeed, 
well-posed ERGMs with additional structure can be well-behaved and are widely used in practice.
We discuss the first point below and the second point in Section \ref{clarification.good}.

The first theoretical results on ill-behaved ERGMs were reported by \citet{St86}, \citet{Jo99}, and \citet{HaJo99}.
\citet{St86} pointed out that the Markov random graphs of \citet{FoSd86},
an important class of ERGMs, 
induce long-range dependence by allowing each edge variable to interact with $2\, (|\verts|-2)$ other edge variables.
The long-range dependence is rooted in the lack of structure of those models:
without additional structure,
it is difficult to constrain the range of interactions.
If strong homogeneity assumptions are imposed on Markov random graphs,
long-range dependence results in strong dependence and model near-degeneracy:
that is,
Markov random graphs concentrate probability mass on a small subset of graphs,
e.g.,
graphs with almost no edges or almost all possible edges \citep{St86}.
\citet{Jo99} and \citet{HaJo99} studied the model near-degeneracy and phase transitions of the triangle model,
a special case of Markov random graphs,
and concluded that the model near-degeneracy of the triangle model is rooted in the lack of structure of the model.
More work on model near-degeneracy can be found in \citet{Sc09b},
\citet{Bu10},
\citet{ChDi11},
\citet{Me17},
and \citet{BhCh18w},
and more work on phase transitions in \citet{ChDi11},
\citet{mukherjee2013phase},
\citet{RaYi13},
\citet{ArRa13},
\citet{yin2016asymptotic},
and \citet{KeYi17};
some related work on phase transitions in the physics literature can be found in \citet{PaNe04,PaNe05}.
\citet{Ha03p} studied the implications of model near-degeneracy in terms of statistical inference,
including the existence of maximum likelihood and Monte Carlo maximum likelihood estimators,
and argued that Monte Carlo maximum likelihood estimators frequently do not exist due to model near-degeneracy,
resulting in computational failure.
\citet{fienberg-2008} investigated these existence issues in more depth by studying the geometry of ERGMs.
The fact that near-degenerate ERGMs have regimes that can be approximated by Bernoulli random graphs in large random graphs was first pointed out by \citet{BaBrSl08,BaBrSl11},
with more work by \citet{ChDi11}, \citet{Me17}, and \citet{BhCh18w}.

Despite the insights gained by these theoretical results,
it is important to keep in mind that these results are limited to simplistic ERGMs that lack structure and have the same number of natural parameters,
regardless of how large $|\verts|$ is \citep[with one exception,][which we discuss in Section \ref{clarification.good}]{Sc09b}.
In particular,
these results do not cover ERGMs with additional structure and ERGMs for which the number of natural parameters increases with $|\verts|$,
which can be better-behaved.
A related limitation of these results is that many real-world settings (e.g., families, classrooms, local groups of insurgents and terrorist cells) involve bounded networks with small numbers of nodes (e.g., 5--50).
In such situations, 
some ``near-degenerate'' specifications may in fact be well-behaved.  
Thus, 
theoretical results suggesting problematic behavior must be considered within the context in which the model is employed \citep[and should checked by simulating graphs from the model, as advised by][]{HuGoHa08}.

\subsubsection{Clarification: Well-posed ERGMs with additional structure can be well-behaved}
\label{clarification.good}

The most important lesson from the theoretical results on simplistic ERGMs is that ERGMs for large networks need additional structure.
The pioneers \citep{St86,StIk90,Jo99,HaJo99} understood full well that the undesirable behavior of simplistic ERGMs is rooted in the lack of structure of those models,
compared with Ising models in physics,
which have additional structure in the form of lattice structure,
and discrete Markov random fields in spatial statistics,
which have additional structure in the form of spatial structure.
To address the lack of structure of simplistic ERGMs,
this early work suggested endowing ERGMs with additional structure.  
For instance, 
\citet{StIk90} introduced Markov random graph models with observed blocks---using categorical covariates to partition the set of nodes into subsets (blocks)---to constrain dependence of Markov random graphs to sets of edge variables within blocks, 
noting that {\em ``Markov models without blocks are unsuitable for large data sets because of the possibility of degeneracy''} \citep[][p.\ 206]{StIk90}.
In the special case of the triangle model,
\citet{Jo99} concluded: {\em ``the random triangle model is explosive; depending on $q$ we get nothing or everything.
The important moral of this is that for any random graph model with transitivity not degenerate in this sense, 
the non-degeneracy relies on the extra,
and perhaps unintended, structure imposed on the graph"} \citep[][p.\ 866]{Jo99}.\footnote{The parameter $q$ mentioned by \citet{Jo99} is equivalent to $q = \exp(\param_2)$,
where $\param_2$ is the triangle parameter of the triangle model stated in Equation \eqref{triangle.model}.} 

In the ERGM framework,
many possible forms of additional structure exist,
including block structure,
multilevel structure,
spatial structure,
or temporal structure.
Additional structure can be used to construct well-behaved ERGMs.
We review ERGMs with additional structure in Section \ref{ergms.additional.structure}.
As a motivating example, however,
we here mention one important instance of ERGMs with additional structure:
curved ERGMs with geometrically weighted model terms \citep{SnPaRoHa04,HuHa04,Hu08}.
Curved ERGMs,
which are curved exponential families in the sense of \citet{Ef75,Ef78},
impose additional structure in the form of nonlinear constraints on the natural parameter space of the exponential family \citep{BN78,Br86}.
The additional structure helps construct better-behaved models.
To demonstrate,
first consider the ill-behaved triangle model studied by \citet{St86}, \citet{Jo99}, \citet{HaJo99}, and others.
The triangle model assumes that the probability mass function of a population graph $\Net_{\verts} \in \{0, 1\}^{\binom{|\verts|}{2}}$ is of the form
\beq
\label{triangle.model}
\M_{\verts,\cnmap(\params,\verts)}(\Net_{\verts} = \net_{\verts})
\propto \exp\left(\param_1 \displaystyle\sum_{i\in\verts < j\in\verts} \edg{i,j} + \param_2 \displaystyle\sum_{i\in\verts < j\in\verts < k\in\verts} \edg{i,j}\, \edg{j,k}\, \edg{i,k}\right),
\eeq
where $\cnmap(\params,\verts) = \params \in \mR^2$,
$\sum_{i\in\verts < j\in\verts} \edg{i,j}$ is the number of edges, 
and\linebreak 
$\sum_{i\in\verts < j\in\verts < k\in\verts} \edg{i,j}\, \edg{j,k}\, \edg{i,k}$ is the number of triangles.
The triangle model with $\param_2 > 0$ rewards transitivity by rewarding triangles.
While transitivity is an important feature of many real-world networks \citep{HpLs70,HpLs72,HpLs76},
the triangle model assumes that the added value of additional triangles does not decrease:
the log odds of a graph with $a$ edges and $b$ triangles relative to a graph with $a$ edges and $b-1$ triangles is 
\beq
\log\dfrac{\M_{\verts,\cnmap(\params,\verts)}(\Net_{\verts} =\; \mbox{graph with $a$ edges and $b$ triangles})}{\M_{\verts,\cnmap(\params,\verts)}(\Net_{\verts} =\; \mbox{graph with $a$ edges and $b-1$ triangles})}
\;=\; \param_2.
\eeq
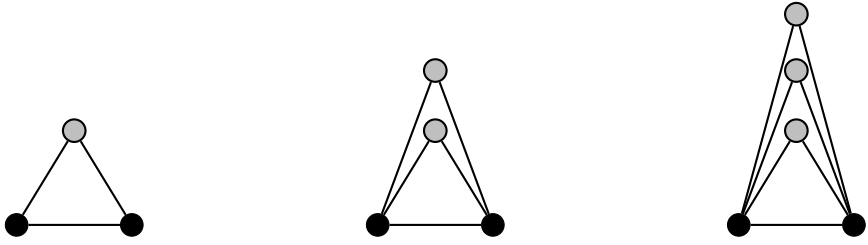
\begin{figure}[t]
\begin{center}
{\small
\begin{tikzpicture}[scale = 0.24]

\begin{scope}[xshift=60cm]
\node[align = flush center] (r_3C) [] {};
\node[female node] (3C_p1) [below left = 0.75 cm of r_3C] {};
\node[female node] (3C_p2) [below right = 0.75 cm of r_3C] {};
\node[main node] (3C_p3) [above  = 0.2 cm of r_3C] {};

\path[draw,thick]
(3C_p1) edge (3C_p2)
(3C_p1) edge (3C_p3)
(3C_p2) edge (3C_p3);

\end{scope}

\begin{scope}[xshift=80cm]
\node[align = flush center] (r_2T) [] {};
\node[female node] (2T_p1) [below left = 0.75 cm of r_2T] {};
\node[female node] (2T_p2) [below right = 0.75 cm of r_2T] {};
\node[main node] (2T_p3) [above = 0.2 cm of r_2T] {};
\node[main node] (2T_p4) [above = 1 cm of r_2T] {};

\path[draw, thick]
(2T_p1) edge (2T_p2)
(2T_p1) edge (2T_p4)
(2T_p1) edge (2T_p3)
(2T_p2) edge (2T_p3)
(2T_p2) edge (2T_p4);

\end{scope}

\begin{scope}[xshift=100cm]
\node[align = flush center] (r_3T) [] {};
\node[female node] (3T_p1) [below left = .75 cm of r_3T] {};
\node[female node] (3T_p2) [below right = .75 cm of r_3T] {};
\node[main node] (3T_p3) [above = 0.2 cm of r_3T] {};
\node[main node] (3T_p4) [above = 1 cm of r_3T] {};
\node[main node] (3T_p5) [above = 1.75 cm of r_3T] {};

\path[draw, thick]
(3T_p1) edge (3T_p2)
(3T_p1) edge (3T_p4)
(3T_p1) edge (3T_p3)
(3T_p2) edge (3T_p3)
(3T_p2) edge (3T_p4)
(3T_p1) edge (3T_p5)
(3T_p2) edge (3T_p5);

\end{scope}

\end{tikzpicture}
}
\caption{\label{esp}
A graphical representation of a connected pair of nodes (represented by two black circles connected by a line),
which has $1$ edgewise shared partner (left), 
$2$ edgewise shared partners (middle), 
and $3$ edgewise shared partners (right).
}
\end{center}
\end{figure}
Note that the log odds does not depend on the number of triangles in the graph.
Indeed,
for each pair of nodes,
each additional triangle contributes the same amount to the log odds of the conditional probability of an edge,
regardless of the number of triangles in which the two nodes are already involved.
This, 
upon reflection, 
is counterintuitive: it makes little sense for the tenth shared partner to carry as much weight as the first.  
Indeed,
the assumption of constant marginal effects implied by the triangle term is not without consequences, 
giving rise to the undesirable behavior described in Section \ref{clarification.theoretical}.
To ensure that the added value of additional triangles decreases,
curved ERGMs with Geometrically Weighted Edgewise Shared Partner ($\gwesp$) terms and other model terms have been developed \citep{SnPaRoHa04,HuHa04,Hu08}.
These models assume that the probability mass function of a population graph $\Net_{\verts} \in \{0, 1\}^{\binom{|\verts|}{2}}$ is of the form
\beq
\M_{\verts,\cnmap(\params,\verts)}(\Net_{\verts} = \net_{\verts})
\propto \exp\left(\eta_1(\params, \verts)\; \displaystyle\sum_{i\in\verts\, <\, j\in\verts} \edg{i,j} +\; \dsum_{m=1}^{|\verts|-2} \eta_{1+m}(\params, \verts)\, \esp_m(\net_{\verts})\right),
\eeq
where the sufficient statistics of the exponential family are the number of edges and the number of connected pairs of nodes with $m$ edgewise shared partners,
$\esp_m(\net_{\verts})$,
and the natural parameters of the exponential family are
\[
\begin{array}{lll}
\eta_{1}(\params, \verts)
\;=\; \param_1\s
\\
\eta_{1+m}(\params, \verts)
\;=\; \param_2\, \exp(\param_3)\, \left[1 - (1 - \exp(-\param_3))^{m}\right],\;\;
m = 1, \dots, |\verts|-2.
\end{array}
\]
A graphical representation of connected pairs of nodes with $1$, $2$, and $3$ edgewise shared partners can be found in Figure \ref{esp}.

\begin{figure}
\begin{center}
\includegraphics[width=\textwidth]{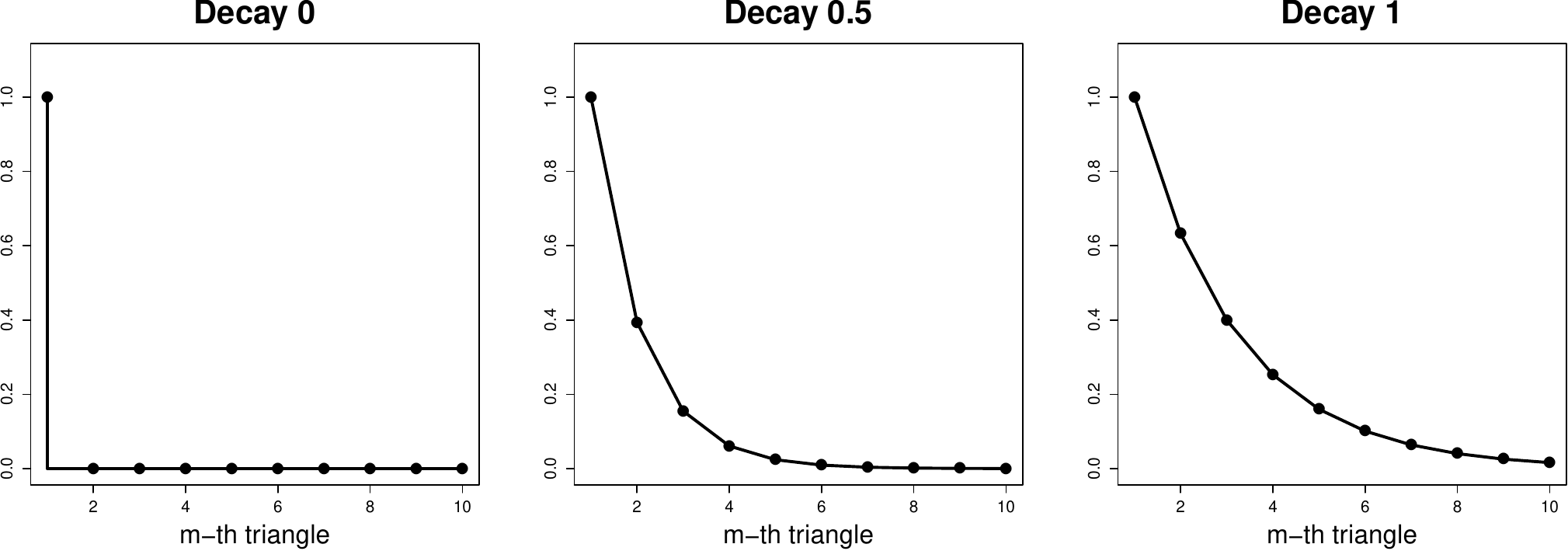}
\caption{\label{cef}
The added value of the $m$-th triangle in terms of $\param_2\, (1 - \exp(-\param_3))^{m-1}$,
$m = 1, 2, \dots$
The plots are based on the so-called base parameter $\param_2=1$ and the so-called decay parameter $\param_3 = 0$ (left), $\param_3 = .5$ (middle), and $\param_3 = 1$ (right).}
\end{center}
\end{figure}

If $\param_2>0$ and $\param_3>0$,
the model rewards triangles,
but ensures that the added value of additional triangles decreases.
To see that,
consider a connected pair of nodes $\{i, j\}$.
The log odds of a graph where $\{i, j\}$ has $m$ shared partners and hence $m$ triangles relative to a graph where $\{i, j\}$ has $m-1$ triangles is,
assuming everything else is the same,
given by
\beq
\label{addedvalue}
\param_2\, (1 - \exp(-\param_3))^{m-1},\;\;
m=1, \dots, |\verts|-2.
\eeq
In other words,
the added value of additional triangles decays at a geometric rate,
provided $\param_2>0$ and $\param_3>0$.
A graphical representation of the added value of additional triangles is shown in Figure \ref{cef}.
Curved ERGMs with geometrically weighted model terms are well-posed as long as $\param_3 \geq 0$;
note that $\param_3 \in [-\log 2,\, 0)$ implies that the added value of the $m$-th triangle either decreases or increases,
depending on the sign of $\param_2$ and whether $m$ is even or odd,
and that $\param_3 \in (-\infty,\, -\log 2)$ implies a form of model near-degeneracy when $|\verts|$ is large \citep{Sc09b}.
In practice,
curved ERGMs with GWESP terms and other geometrically weighted model terms have turned out to be well-behaved in a wide range of settings;
selected applications can be found in \citet{SnPaRoHa04}, \citet{HuHa04}, \citet{Hu08}, \citet{HuGoHa08}, \citet{GoKiMo2009}, \citet{GiHa06}, \citet{HaGi09}, \citet{KoRoPa09}, \citet{SiHaLa11}, \citet{Su12}, \citet{Rolletal13}, \citet{Wa13}, \citet{AlBa15}, \citet{brain17}, \citet{gondal2018duality}, and \citet{StScBoMo18}.
We apply curved ERGMs to human brain network data in Section \ref{application},
illustrating how curved ERGMs can provide advantages over both Bernoulli random graphs and latent space models in some applications.

In addition to their enhanced stability,
there exist consistency results for curved ERGMs with GWESP terms and other curved ERGMs.
We review consistency results for curved ERGMs in Section \ref{ergm.consistency}.
It is worth noting that the results of \citet{ChDi11} and others do not preclude consistency results for curved ERGMs:
the results of \citet{ChDi11} are limited to ERGMs which are canonical exponential families with a fixed number of natural parameters of the form $\eta_i(\params) = \param_i$,
and those results do not cover curved exponential families with an increasing number of natural parameters subject to nonlinear constraints,
such as curved ERGMs with GWESP terms and $1 + |\verts|-2$ natural parameters.

\subsection{Question II: Must ERGMs be projective for well-posed inference?}
\label{concern.nonprojective}

\subsubsection{Overview and history}
\label{projectivity.bad}

A second question that has been raised is whether ERGMs must satisfy a property called projectivity for statistical inference to be well-posed.
This question arises from the observation that many ERGMs are not projective, 
as first pointed out by \citet{La08} and \citet{Sn10}.
Necessary and sufficient conditions for projectivity of ERGMs with size-invariant natural parameters of fixed dimension were provided by \citet{ShRi11};
some follow-up work can be found in \citet{LaRiSa17}.
Other notable work related to the topic of projectivity---outside of the ERGM framework---can be found in,
e.g.,
\citet{CaFo17}, 
\citet{crane2018probabilistic},
and \citet{CrDe15}.

In the ERGM framework, 
projectivity may be defined as follows.  
Consider a subgraph $\net_{\verts'}$ of a population graph $\net_{\verts}$ induced by a subset of nodes $\verts' \subset \verts$,
i.e.,
the subgraph with the set of nodes $\verts'$ and all edges among nodes in $\verts'$ contained in the population graph $\net_{\verts}$.
An ERGM is called projective if
\beq
\cnmap(\params,\verts')
\eqs \params
\;\; \mbox{ for all }\;\; \params\in\{\params\in\mR^p: \psi(\params, \verts') < \infty\}\;\; \mbox{ and all }\;\; \verts' \subseteq \verts
\eeq
and
\beq
\M_{\verts',\params}(\Net_{\verts'} = \net_{\verts'})
= \M_{\verts,\params}(\Net_{\verts'} = \net_{\verts'}, \Net_{\verts \setminus \verts'} \in \netspace_{\verts \setminus \verts'}) \mbox{ for all  } \params\in\paramf \mbox{ and all } \verts'\subseteq\verts,
\eeq
where $\net_{\verts \setminus \verts'} \in \netset_{\verts \setminus \verts'}$ denotes the subset of possible edges of the population graph which are not contained in the subgraph induced by $\verts'$.
The notion of projectivity above was advanced by \citet{ShRi11},
who focused on ERGMs with counting measure as reference measure,
although \citeauthor{ShRi11} considered more general reference measures in a supplement.
We consider here the more general definition,
covering all finite reference measures.
It is worth noting that there are weaker forms of projectivity in the ERGM framework,
e.g., 
conditional projectivity \citep{Sn10} and block projectivity \citep{ScHa13}.
Here, 
we restrict attention to the definition of projectivity mentioned above.

Projectivity is a form of closure under marginalization,
and implies that the same parameters govern the population graph and the marginal distributions of all of its subgraphs.
\alert{While closure under marginalization is convenient on mathematical grounds,
it embodies the strong assumption that no subgraph of the population graph is affected by its embeddedness in the population graph.
For example,
consider $3$ people attending a party of $30$.
To be projective,
the probabilities of interactions among those $3$ people must neither be affected by interactions of those $3$ with the $27$ others nor by the interactions among the $27$ others.
In other words,
the probabilities of interactions among the $3$ people must be the same,
regardless of whether $27$ other people attend the party.  
Such an assumption may not be satisfied by many networks (human or otherwise). 
}
Not surprisingly, 
many ERGMs are not projective,
including some of the simplest and most classic random graph models.

For instance,
sparse Bernoulli$(\pi_{|\verts|})$ random graphs with size-dependent edge probabilities $\pi_{|\verts|}$ \citep{Gi59,ErRe59,ErRe60} are not projective, 
because the edge probabilities $\pi_{|\verts|}$ decrease with the size $|\verts|$ of $\verts$.
Sparse Bernoulli$(\pi_{|\verts|})$ random graphs assume that the edge variables $\Edge{i,j}$ are independent Bernoulli$(\pi_{|\verts|})$ random variables,
and that the expected number of edges $\binom{|\verts|}{2}\, \pi_{|\verts|}$ grows slower than $\binom{|\verts|}{2}$,
which implies that $\pi_{|\verts|}$ must decrease with $|\verts|$.
Consider the parameterization $\pi_{|\verts|} = \logit^{-1}(\theta - \log |\verts|)$ ($\param\in\mR$),
which implies
\beq
\begin{array}{lll}
\label{bernoulli.pmf}
\M_{\verts,\eta(\param,\verts)}(\Net_{\verts} = \net_{\verts})
&=& \displaystyle\prod_{i\in\verts < j\in\verts} \pi_{|\verts|}^{y_{i,j}}\, (1-\pi_{|\verts|})^{1-y_{i,j}} \s
\\
&\propto& \exp\left(\eta(\param, \verts) \displaystyle\sum_{i\in\verts < j\in\verts} \edg{i,j}\right) \mref(\net_{\verts}),
\end{array}
\eeq
where 
\beq
\eta(\param,\verts')
\eqs \param
\;\; \mbox{ for all }\;\; \param\in\mR\;\; \mbox{ and all }\;\; \verts' \subseteq \verts
\eeq
and the offset $\log |\verts|$ has been absorbed into the reference measure:
\beq
\label{nu}
\mref(\net_{\verts}) \;=\; 
\begin{cases}
\exp\left(- \log |\verts| \displaystyle\sum_{i\in\verts < j\in\verts} \edg{i,j}\right) & \mbox{ if } \net_{\verts} \in \{0, 1\}^{\binom{|\verts|}{2}}\s
\\
0 & \mbox{ otherwise.}
\end{cases}
\eeq
Sparse Bernoulli$(\pi_{|\verts|})$ random graphs are not projective,
because $|\verts'| \neq |\verts|$ implies $\pi_{|\verts'|} \neq \pi_{|\verts|}$,
so
\begin{align*}
\prod_{i\in\verts' < j\in\verts'} \pi_{|\verts'|}^{y_{i,j}}\, (1-\pi_{|\verts'|})^{1-y_{i,j}} &\ne \prod_{i\in\verts' < j\in\verts'} \pi_{|\verts|}^{y_{i,j}}\, (1-\pi_{|\verts|})^{1-y_{i,j}}
\end{align*}
and hence
\begin{align*}
\vphantom{\sum_{\net_{\verts \setminus \verts'}\; \in\; \netset_{\verts \setminus \verts'}}}\M_{\verts',\param}(\Net_{\verts'} = \net_{\verts'}) \;\;\ne\;\; \M_{\verts,\param}(\Net_{\verts'} = \net_{\verts'},\; \Net_{\verts \setminus \verts'} \in \netspace_{\verts \setminus \verts'}).
\label{bernoulli.nonprojective}
\end{align*}
\vspace{-.5cm}

\noindent
It is worth noting that,
despite the lack of projectivity,
sparse Bernoulli$(\pi_{|\verts|})$ random graphs have meaningful asymptotic behavior:
e.g.,
the expected number of edges of each node tends to the constant $\exp(\theta)$ as $|\verts|\rightarrow\infty$ \citep{KrHaMo11}.
Thus,
lack of projectivity does not rule out meaningful asymptotic behavior.
Indeed, 
sparse Bernoulli$(\pi_{|\verts|})$ random graphs have many interesting asymptotic properties and have been studied in random graph theory since the seminal work of \citet{Gi59} and \citet{ErRe59,ErRe60}:
see,
e.g.,
the classic monograph of \citet{Bo85} and the more recent books of \citet{JaLuRu00} and \citet{FrKa16}.  
Moreover, 
the sparsity-inducing reference measure \eqref{nu} can be shown to arise from a simple stochastic process with an attractive interpretation \citep{butts:jms:2018}, 
providing a substantive motivation for its use.

The above suggests that projectivity is too restrictive to be a basis for building plausible models of complex and dependent network data. 
Nonetheless, 
projectivity is convenient, 
in that it provides a simple basis for extending an ERGM from a subgraph to the population graph.
This raises the question of whether, 
in the absence of projectivity, 
population probability models can be inferred from a subgraph of the population graph.

\subsubsection{Clarification: The likelihood function is not affected by lack of projectivity}
\label{clarification.likelihood}

While many ERGMs are not projective, 
lack of projectivity does not preclude likelihood-based inference,
for at least two reasons.
First,
the likelihood function is not affected by lack of projectivity.
Second,
lack of projectivity does not imply that likelihood-based estimators are inconsistent.
In fact,
consistency results for likelihood-based estimators of non-projective ERGMs do exist.
Taken together,
these two points imply that likelihood-based inference for well-posed ERGMs is possible despite lack of projectivity.
We discuss the first point below and the second point in Section \ref{clarification.consistency}.

The first important point is that the likelihood function is not affected by lack of projectivity.
Consider the motivating example of \citet[p.\ 509]{ShRi11}:
there is a finite population of nodes $\verts$ and a population graph $\net_{\verts}$ is generated by a population probability model $\M_{\verts,\cnmap(\params,\verts)}(\Net_{\verts} = \net_{\verts})$.  
This motivating example is representative of other ERGM applications:
e.g.,
in the human brain network example in Section \ref{application},
the population of nodes $\verts$ corresponds to 56 regions of the human brain.

To derive the likelihood function in the motivating example,
it is imperative to separate the process that generates the population graph from the process that generates observations of edges in the population graph.
We follow here the principled approach of \citet{Fi22} and \citet{Ru76} to likelihood-based inference in complete- and incomplete-data scenarios,
which was adapted to ERGMs by \citet{GiHa06}, \citet{HaGi09}, and \citet{KoRoPa09}.
If the whole population graph $\net_{\verts}$ is observed,
the likelihood function is
\beq
\label{likelihood.complete}
\lik(\params;\; \net_{\verts})
\;\propto\; \M_{\verts,\cnmap(\params,\verts)}(\Net_{\verts} = \net_{\verts}).
\eeq
If a subgraph $\net_{\verts'}$ of $\net_{\verts}$ induced by a subset of nodes $\verts' \subset \verts$ is observed,
generated by an incomplete-data generating process that is ignorable in the sense of \citet{Ru76},
the likelihood function is
\beq
\label{likelihood.incomplete}
\lik(\params;\; \net_{\verts'})
\;\propto\; \M_{\verts,\cnmap(\params,\verts)}(\Net_{\verts'} = \net_{\verts'},\; \Net_{\verts \setminus \verts'} \in \netspace_{\verts \setminus \verts'}).
\eeq
In other words,
the likelihood function can be obtained by summing the population probability mass function $\M_{\verts,\cnmap(\params,\verts)}(\Net_{\verts} = \net_{\verts})$ with respect to the unobserved edges.

While deriving the likelihood function \eqref{likelihood.incomplete} is trivial from a mathematical point of view---although computing it may not be trivial, as discussed below---the form of the likelihood function has important statistical implications.
First,
as \citet{ShRi11} point out, it is problematic to base inference on $\M_{\verts',\cnmap(\params,\verts')}(\Net_{\verts'} = \net_{\verts'})$,
because $\M_{\verts',\cnmap(\params,\verts')}(\Net_{\verts'} = \net_{\verts'})$ may not be relatable to $\M_{\verts,\cnmap(\params,\verts)}(\Net_{\verts} = \net_{\verts})$ when the model is not projective.
It is therefore comforting to know that likelihood-based inference is not based on $\M_{\verts',\cnmap(\params,\verts')}(\Net_{\verts'} = \net_{\verts'})$,
but is based on the marginal probability mass function $\M_{\verts,\cnmap(\params,\verts)}(\Net_{\verts'} = \net_{\verts'},\; \Net_{\verts \setminus \verts'} \in \netspace_{\verts \setminus \verts'})$ induced by $\M_{\verts,\cnmap(\params,\verts)}(\Net_{\verts} = \net_{\verts})$.
Indeed,
the marginal probability mass function is related to $\M_{\verts,\cnmap(\params,\verts)}(\Net_{\verts} = \net_{\verts})$ by marginalization,
regardless of whether the model is projective.
As a result,
the likelihood function \eqref{likelihood.incomplete} is not affected by lack of projectivity.  
Projectivity does facilitate the evaluation of the likelihood function,
but the likelihood function does not require it.

\s

\paragraph*{Statistical implications.}

The results of \citet{ShRi11} underscore the importance of likelihood-based inference:
statistical inference for ERGMs should be based on the likelihood function \eqref{likelihood.incomplete},
which is not affected by lack of projectivity.
Indeed,
since the 1990s,
ERGM experts have known that non-likelihood-based inference is problematic.
\citet{DaDa93,DaDa99}, 
\citet{CjDkDp98,CoDaDa02}, 
\citet{LuSni2007}, and \citet{DuGiHa09} compared likelihood-based and non-likelihood-based estimators by using exact computations,
based on complete enumeration of all possible graphs of small sizes,
along with simulation studies and data analyses for graphs of larger sizes.
All of them concluded that non-likelihood-based inference---in particular pseudo-likelihood-based inference \citep{StIk90}---tends to be inferior to likelihood-based inference.
We review likelihood-based inference given incomplete data in Section \ref{sec:likelihood}.

\s

\paragraph*{Computational implications.}

While projectivity is not necessary,
it is convenient for the purpose of evaluating the likelihood function \eqref{likelihood.incomplete}.
If the model is projective,
summing $\M_{\verts,\cnmap(\params,\verts)}(\Net_{\verts} = \net_{\verts})$ with respect to the unobserved edges by using computational methods is unnecessary,
because the likelihood function \eqref{likelihood.incomplete} reduces to $\M_{\verts',\cnmap(\params,\verts')}(\Net_{\verts'} = \net_{\verts'})$.
Otherwise,
one needs to sum $\M_{\verts,\cnmap(\params,\verts)}(\Net_{\verts} = \net_{\verts})$ with respect to the unobserved edges by using computational methods,
either exactly or approximately, 
by using Markov chain Monte Carlo methods \citep{GiHa06,HaGi09,KoRoPa09}.
These computational challenges are the same as in other discrete exponential-family models for dependent random variables,
such as discrete Markov random fields in spatial statistics \citep{Bj74,Cr93,St99}.
In many applications of ERGMs,
the required computations are feasible,
because either the population of interest is not too large or the population has additional structure that facilitates computations.  

An example of small populations are the human brain networks used in Section \ref{application}:
each of the 108 human brain networks has 56 nodes and is therefore small enough to approximate the likelihood function \eqref{likelihood.incomplete} by using Markov chain Monte Carlo methods when network data are sampled or missing.

Examples of populations with additional structure are populations with block structure, multilevel structure, spatial structure, and temporal structure.
Some ERGMs with additional structure have factorization properties.
Factorization properties facilitate likelihood-based computations by allowing to break down likelihood functions into parts,
and the parts may be computed by using parallel computing on multi-core computers and computing clusters.
We review ERGMs with additional structure,
including ERGMs with factorization properties,
in Section \ref{ergms.additional.structure}.

Such likelihood-based methods are implemented in many ERGM-related statistical software packages---including 20 ERGM-related \texttt{R} packages (listed in footnote \ref{ergm.r} in Section \ref{sec:introduction}) and the program \texttt{pnet} \citep{pnet}---and the computational burden has not prevented network scientists from applying ERGMs to a substantial number of real-world problems (with some examples given in Section \ref{sec:introduction}).

\subsubsection{Clarification: Consistency results for likelihood-based estimators of non-projective ERGMs do exist}
\label{clarification.consistency}

The second important point regarding projectivity is that lack of projectivity does not imply that estimators of $\params$ are inconsistent.
\citet{ShRi11} showed that projectivity is a {\em sufficient} condition for consistency of maximum likelihood estimators for size-invariant natural parameters of fixed dimension.
However,
projectivity is not a \emph{necessary} condition for consistency of maximum likelihood estimators:
there do exist consistency results for maximum likelihood estimators of non-projective ERGMs.
We review them in Section \ref{ergm.consistency}.

\subsubsection{Clarification: Typically, ERGMs are not applied to large populations, unless there is additional structure}
\label{clarification.largepopulations}

An implicit assumption,
underlying both questions described above,
is that ERGMs are applied to large populations without additional structure.
However,
in most applications to date,
ERGMs have been applied to either small populations or large populations with additional structure, 
as pointed out in Section \ref{clarification.likelihood}.
Statistical theory should take advantage of additional structure rather than working under the assumption that ERGMs are applied without it.
Advances in the development of concentration inequalities \citep[e.g.,][]{Talagrand96,BoLuMa13} enable statistical theory to do so.
We review an example in Section \ref{consistency.super}.

\section{ERGMs with additional structure}
\label{ergms.additional.structure}

Well-posed ERGMs with additional structure address the issues of simplistic ERGMs and therefore have an important place in the ERGM framework,
as explained in Section \ref{clarification.good}.
Here,
``additional structure'' is understood as additional mathematical structure imposed on\s
\bi
\item[(a)] the support or reference measure of ERGMs;
\item[(b)] the dependence structure of ERGMs;
\item[(c)] the parameter space of ERGMs;
\item[(d)] combinations of (a), (b), and (c).
\ei
\s

For example,
additional structure may come in the form of constraints on the number of edges or other functions of the random graph;
constraints on the parameter space of ERGMs,
as imposed by curved ERGMs;
or constraints on the dependence structure of ERGMs imposed by block structure,
multilevel structure,
spatial structure,
or temporal structure.

We review here some of the more established classes of ERGMs with additional structure.
Other classes,
not discussed here,
can be found in \citet[][ERGMs with latent graphs]{WyChBi08}, 
\citet[][support-restricted ERGMs]{karwa2016dergms},
\citet[][nonparametric ERGMs with constraints]{ThKa17},
and \citet[][constraining variances of sufficient statistics]{FeHa17}.

\subsection{Dyad-independent ERGMs}
\label{dyadindependence.ergms}

A simple form of additional structure comes in the form of constraints on the dependence structure of ERGMs:
in the simplest case,
one can assume that all edge variables $\Edge{i,j}$ in undirected random graphs are independent or all dyads $(\Edge{i,j}, \Edge{j,i})$ in directed random graphs are independent.

In the undirected case,
assuming edge variables $\Edge{i,j}$ are independent,
the population probability mass function satisfies the following factorization property:
\beq
\label{betamodel}
\begin{array}{lll}
\M_{\verts,\cnmap(\params,\verts)}(\Net_{\verts} = \net_{\verts})
&=& \displaystyle\prod_{i\in\verts\, <\, j\in\verts} \M_{\{i,j\},\cnmap(\params,\verts)}(\Edge{i,j} = \edg{i,j}).
\end{array}
\eeq
In the directed case,
assuming dyads $(\Edge{i,j}, \Edge{j,i})$ are independent,
the population probability mass function satisfies the following factorization property:
\beq
\label{p1model}
\begin{array}{lll}
\M_{\verts,\cnmap(\params,\verts)}(\Net_{\verts} = \net_{\verts})
&=& \displaystyle\prod_{i\in\verts\, <\, j\in\verts} \M_{\{i,j\},\cnmap(\params,\verts)}(\Edge{i,j} = \edg{i,j},\, \Edge{j,i} = \edg{j,i}).
\end{array}
\eeq
Classic examples are Bernoulli random graphs and $\beta$-models for undirected random graphs \citep[][]{Gi59,ErRe59,ErRe60,DiChSl11,RiPeFi13,karwa2016inference} and $p_1$-models for directed random graphs \citep{HpLs81,YaZhQi15,YaLeZh11,YaWaQi16,Yaetal18}.
These models can capture heterogeneity in the propensities of nodes to form edges:
e.g.,
$\beta$-models assume that the edge variables $\Edge{i,j}$ are independent Bernoulli$(\pi_{i,j})$ random variables with node-dependent edge probabilities $\denspar_{i,j} = \ilogit(\param_i + \param_j)$ ($\param_i\in\mR$, $\param_j\in\mR$),
which are equivalent to ERGMs with probability mass functions of the form
\beq
\begin{array}{llllll}
\M_{\verts,\cnmap(\params,\verts)}(\Net_{\verts} = \net_{\verts})
\;\propto\; \displaystyle\prod_{i\in\verts\, <\, j\in\verts} \exp\left(\eta_{i,j}(\params, \verts)\; \edg{i,j}\right),
\end{array}
\eeq
where $\cnmap(\params, \verts)$ can be viewed as a vector with elements
\beq
\eta_{i,j}(\params, \verts) \;\;=\;\; \param_i + \param_j,
\;\;\;\; i \in \verts,\;\;\;\; j \in \verts.
\eeq
The parameters $\param_i$ and $\param_j$ can be interpreted as the propensities of nodes $i \in \verts$ and $j \in \verts$ to form edges,
respectively.
These propensities can vary from node to node,
so the model can capture heterogeneity in the propensities of nodes to form edges.
In addition,
when the edges in the population graph are directed and the population graph is generated by $p_1$-models \citep{HpLs81},
such models can capture reciprocity.
Reciprocity refers to the tendency of nodes $i\in\verts$ and $j\in\verts$ to reciprocate edges, 
i.e.,
$\Edge{i,j}=1$ and $\Edge{j,i}=1$ are observed more frequently than would be expected when the edges $\Edge{i,j}$ and $\Edge{j,i}$ were independent.
Note that reciprocity induces dependence between edge variables $\Edge{i,j}$ and $\Edge{j,i}$,
but leaves dyads $(\Edge{i,j}, \Edge{j,i})$ independent.

Dyad-independent ERGMs constrain the range of dependence and therefore do not have the undesirable properties of the triangle model and other simplistic ERGMs,
which are rooted in the lack of structure and the strong dependence induced by those models,
as explained in Sections \ref{clarification.theoretical} and \ref{clarification.good}.
In addition,
the factorization properties of probability mass functions \eqref{betamodel} and \eqref{p1model} have computational advantages,
facilitating the evaluation of likelihood functions.

\subsection{Curved ERGMs}
\label{curved.ergms}

While dyad-independent ERGMs do not have the undesirable properties of the triangle model and other simplistic ERGMs,
such models do not capture model dependencies among edge variables other than reciprocity in directed random graphs.
To model transitivity and other network phenomena inducing dependence,
ERGMs with less restrictive forms of additional structure have to be considered.
Curved ERGMs are an important example.

Curved ERGMs were developed by \citet{SnPaRoHa04} and \citet{HuHa04} to address the flaws of simplistic ERGMs lacking structure,
such as the triangle model \citep[see, e.g.,][]{SnPaRoHa04,HuHa04,Hu08,HuGoHa08,GoKiMo2009,RoPaWa09}.
Curved ERGMs impose additional structure in the form of constraints on the natural parameter space of the exponential family \citep{BN78,Br86} and are curved exponential families in the sense of \citet{Ef75,Ef78}.
We presented one example of curved ERGMs in Section \ref{clarification.good}:
curved ERGMs with edge and GWESP terms,
which ensure that the added value of additional triangles decreases,
in contrast to the triangle model.
To describe a large class of curved ERGMs with geometrically weighted model terms,
consider ERGMs with probability mass functions of the form
\beq
\M_{\verts,\cnmap(\params, \verts)}(\Net_{\verts} = \net_{\verts})
\;\propto\; \exp\left(\dsum_{m=1}^{|\verts|-2} \eta_{m}(\params, \verts)\, \genstat_m(\attrs_{\verts}, \net_{\verts}) + \dots\right),
\eeq
where the dots refer to additional model terms,
such as edge terms.
Here,
the sufficient statistics $\genstat_m(\attrs_{\verts}, \net_{\verts})$ count the number of units of a specified type with $m$ subgraph configurations of a specified type.
The units may refer to nodes;
unconnected pairs of nodes;
or connected pairs of nodes,
as in GWESP terms.
The subgraph configurations may refer to shared partners,
as in GWESP terms,
but many other subgraph configurations are possible.
Some examples can be found in \citet{HuGoHa08}.
In addition, 
the sufficient statistic may depend on the attributes $\attrs_{\verts}$ of the population of nodes $\verts$.
The natural parameters are of the form
\beq
\label{geos}
\begin{array}{lll}
\eta_m(\params, \verts)
\;=\; \param_1\, \exp(\param_2)\, \left[1 - (1 - \exp(-\param_2))^{m}\right],\;\;
m = 1, 2, \dots
\end{array}
\eeq

The additional structure imposed by these curved ERGMs comes in the form of nonlinear constraints on the natural parameter space of the exponential family.
In the example used in Section \ref{clarification.good},
curved ERGMs with edge and GWESP terms,
the natural parameter space is $\mR^{1 + |\verts| - 2}$,
and the curved ERGM imposes nonlinear constraints on $\mR^{1 + |\verts| - 2}$.
If $\param_1>0$ and $\param_2>0$,
these nonlinear constraints ensure that the added value of additional subgraph configurations of the specified type decreases.
We can see that by examining the log odds of a single unit, e.g., a connected pair of nodes $\{i,j\}$. 
The log odds of a graph where the unit has $m$ configurations of the specified type relative to a graph where it has $m-1$ configurations is,
assuming everything else is the same,
given by
\[
\param_1\, (1 - \exp(-\param_2))^{m-1},\;\;
m=1, \dots, |\verts|-2.
\]
In other words,
the added value of additional configurations of the specified type decreases at a geometric rate.
Figure \ref{cef} in Section \ref{clarification.good} demonstrates it in the special case of GWESP,
where the added value of additional triangles decreases at a geometric rate.
\hide{
\[
\begin{array}{cccccccccccccccccccc}
\underbrace{\param_1}
&>& \underbrace{\param_1\, (1 - \exp(-\param_2))}
&>& \underbrace{\param_1\, (1 - \exp(-\param_2))^2}
&> \dots\\
\mbox{\em added value of}
&& \mbox{\em added value of}
&& \mbox{\em added value of}
\\
\mbox{\em $m=1$st configuration}
&&\mbox{\em $m=2$nd configuration}
&&\mbox{\em $m=3$rd configuration}
\end{array}
\]
}

The main ERGM-related statistical software packages,
including the 19 ERGM-related \texttt{R} packages mentioned in Section \ref{clarification.likelihood} and the program \texttt{pnet} \citep{pnet},
implement many geometrically weighted model terms and related model terms \citep[e.g.,][]{Bu08,RoPaWa09}.
In practice,
curved ERGMs with geometrically weighted terms have been found useful:
selected examples are cited in Section \ref{clarification.likelihood}.
We illustrate that in Section \ref{application} that curved ERGMs with edge, GWESP, and other terms can outperform both Bernoulli random graphs and latent space models.

\subsection{ERGMs with block structure}
\label{block.ergms}

A simple form of additional structure is block structure,
which is popular in the literature on stochastic block models \citep[e.g.,][]{NkSt01,BiCh09,RoChYu11}.
In the simplest case,
a block structure corresponds to a partition of a population of nodes $\verts$ into $K$ subpopulations $\mA_1, \dots, \mA_K$,
called blocks.
While stochastic block models assume that edges within and between blocks are independent conditional on the block structure,
ERGMs with block structure allow edges to be dependent within and between blocks.
As a result,
ERGMs with block structure can be viewed as generalizations of stochastic block models.
As in stochastic block models,
the block structure may be observed or unobserved.

We present here two classes of ERGMs with block structure:
one class of ERGMs that exploits block structure to constrain the range of dependence and another class of ERGMs that exploits block structure to capture unobserved heterogeneity.

\subsubsection{Constraining the range of dependence}
\label{localdependence}

ERGMs that exploit observed block structure to constrain the range of dependence were introduced by \citet{StIk90} in the context of Markov random graphs,
using categorical covariates to partition a set of nodes into blocks.
A more general class,
with observed and unobserved blocks, 
was developed by \citet{ScHa13} and \citet{Sc17}.
For simplicity,
we focus henceforth on observed block structure.

Given observed block structure,
these models are characterized by the factorization condition
\beq
\label{localdependence.ergm}
\begin{array}{lll}
\M_{\verts,\cnmap(\params,\verts)}(\Net_{\verts} = \net_{\verts})
&=& \displaystyle\prod_{k=1}^K \prod_{l=1}^{K-1} \M_{\{\verts_k,\verts_l\},\cnmap(\params,\verts)}(\Net_{\verts_k,\verts_l} = \net_{\verts_k,\verts_l}),
\end{array}
\eeq
where $\Net_{\verts_k,\verts_l}$ denotes the set of edge variables corresponding to possible edges between nodes in subpopulation $\verts_k$ and nodes in subpopulation $\verts_l$.
To constrain the range of dependence to blocks,
these models assume that between-block edges are independent,
\[
\begin{array}{lll}
\M_{\{\verts_k,\verts_l\},\cnmap(\params,\verts)}(\Net_{\verts_k,\verts_l} = \net_{\verts_k,\verts_l})
&=& \displaystyle\prod_{i\in\verts_k,\, j\in\verts_l} \M_{\{i,j\},\cnmap(\params,\verts)}(\Edge{i,j} = \edg{i,j}),
\end{array}
\]
whereas within-block edges may be dependent:
\[
\begin{array}{lll}
\M_{\{\verts_k,\verts_k\},\cnmap(\params,\verts)}(\Net_{\verts_k,\verts_k} = \net_{\verts_k,\verts_k})
&\neq& \displaystyle\prod_{i\in\verts_k\, <\, j\in\verts_k} \M_{\{i,j\},\cnmap(\params,\verts)}(\Edge{i,j} = \edg{i,j}).
\end{array}
\]
If ERGMs are used as within- and between-block models,
then the population probability model is an ERGM:
an ERGM with additional structure in the form of block structure and local dependence within blocks.
A concrete example is given by block-dependent curved ERGMs with block-dependent edge and GWESP terms,
as described in Section \ref{consistency.super}.

Such ERGMs,
exploiting block structure to constrain the range of dependence,
address the lack of structure of simplistic ERGMs and have at least three advantages.
First, 
local dependence often makes substantive sense,
because many real-world networks are local in nature \citep[e.g.,][]{Hg50,WsFk94,PpRg02}.
Second,
these models retain the main advantage of ERGMs,
the flexibility to model a wide range of network features,
because the within-block ERGMs can model a wide range of network features within blocks.
At the same time,
the models address the main disadvantage of simplistic ERGMs,
the lack of structure,
by using block structure to constrain the range of dependence.
As long as the blocks are not too large,
the overall dependence induced by the model is weak and the model does not suffer from model near-degeneracy,
which results from strong long-range dependence,
as discussed in Sections \ref{clarification.theoretical} and \ref{clarification.good}.
Third,
these models have statistical advantages,
because weak dependence can be exploited to derive concentration results,
which in turn can be used to establish consistency results for likelihood-based estimators.
We discuss them in Section \ref{ergm.consistency}.
Last,
but not least,
ERGMs with block structure have computational advantages,
because the factorization of probability mass function \eqref{localdependence.ergm} facilitates the computation of likelihood functions.

\subsubsection{Capturing unobserved heterogeneity}
 
We discuss here two classes of ERGMs that take advantage of block structure to capture unobserved heterogeneity.

The ERGMs of \citet{Ko09c} use block-dependent edge terms to capture unobserved heterogeneity in the propensities of nodes to form edges along with alternating $k$-triangle terms to capture transitivity;
note that alternating $k$-triangle terms are related to GWESP terms,
as explained by \citet{Hu08}.
\citet{Waetal18} used ERGMs with local dependence within blocks \citep{ScHa13} to capture heterogeneity among communities in terms of the propensities to form edges as well as other network features.

\citet{ToMu14} developed mixtures of ERGMs for network data that are collected by sampling nodes from a population of nodes (called egos),
recording which nodes are connected to egos (called alters),
and recording which of the alters are connected.
\citeauthor{ToMu14} assumed that these ego-centric networks were generated by a finite mixture of ERGMs.
The resulting mixture of ERGMs can be viewed as a model with block structure,
where each ego belongs to one block,
each block has a block-dependent ERGM,
and the ego-centric networks of all egos in the same block are generated by the same block-dependent ERGM.
It is worth noting though that the resulting model is a model of ego-centric networks rather than the population graph,
and each possible edge shows up in two ego-centric networks and is hence governed by two block-dependent ERGMs \citep[see][]{ToMu14}.

\subsection{ERGMs with multilevel structure}
\label{multilevel.ergms}

There is a large and growing body of work on multilevel network data and models \citep[e.g.,][]{Lu2003,Wa13,ZaLo15,Lo16,slaughter2016multilevel,hollway2016multilevel,multilevelnetwork,wang2016social,brailly2016embeddedness,meredith2017information,hollway2017multilevel,gondal2018duality}.

Multilevel network data are network data with additional structure in the form of ``nesting'' of nodes within higher-order units.
A simple form of multilevel structure is block structure,
as described in Section \ref{block.ergms}.
Examples are populations of students partitioned into school classes,
armed forces partitioned into units of armed forces, 
and workforces in companies partitioned into departments.
In all of these examples,
the block structure can be observed:
e.g.,
it can be observed who belongs to which unit of armed forces.
More general forms of multilevel structure exist.
There can be three or more levels of multilevel structure, 
e.g.,
in universities,
faculty members (level-1 units) are nested in departments (level-2 units),
departments are nested in schools (level-3 units),
and schools are nested in universities (level-4 units).

To demonstrate multilevel ERGMs,
consider two-level networks,
with nodes (level-1 units) nested in sets of nodes $\verts_1, \dots, \verts_K$ (level-2 units).
An example of two-level networks is given by the human brain networks used in Section \ref{application},
where the level-1 units correspond to 56 regions of the brain and the level-2 units correspond to 108 brains.
A simple form of two-level ERGM assumes that
\beq
\label{twolevel.ergm}
\begin{array}{lll}
\M_{\verts,\cnmap(\params,\verts)}(\Net_{\verts} = \net_{\verts})
&=& \displaystyle\prod_{k=1}^K \M_{\{\verts_k,\verts_k\},\cnmap(\params,\verts)}(\Net_{\verts_k,\verts_k} = \net_{\verts_k,\verts_k}).
\end{array}
\eeq
An ERGM with two-level structure \eqref{twolevel.ergm} is a special case of an ERGM with block structure \eqref{localdependence.ergm} when edges between subpopulations do not exist with probability $1$.
A specific example of a two-level ERGM with edge,
GWESP,
and other terms can be found in Section \ref{application}.
Well-posed ERGMs with multilevel structure share the advantages of ERGMs with block structure,
as discussed in Section \ref{block.ergms}.

\subsection{ERGMs with spatial structure}
\label{spatial.ergms}

In some applications of ERGMs,
the population of nodes is embedded in a space.
The space in question may be a geographical space \citep{butts.acton:ch:2011} or a social space \citep{mcpherson:asr:1983}, 
constructed from observed attributes of population members (e.g., race).
Spatial structure can be exploited to construct more realistic and better-behaved ERGMs:
e.g.,
researchers may hypothesize that the probability of an edge decreases as the distance between population members increases.
To test such hypotheses and estimate the strength of the effect of distance on the population graph,
ERGMs can be used,
with model terms that are functions of the distances between population members.
Such ERGMs can induce sparsity by penalizing edges between pairs of nodes that are separated by large distances,
and can help control the dependence of edge variables,
resulting in better-behaved ERGMs.

To give a simple example,
assume that the spatial locations of nodes are observed,
e.g., when positions are based on social or geographical attributes of nodes.
(We do not consider unobserved spatial structure,
but note that ERGMs with unobserved spatial structure can be viewed as generalizations of latent space models \citep{HpRaHm01}.)
A simple example of an ERGM with observed spatial structure is an ERGM with probability mass function
\beq
\M_{\verts,\cnmap(\params, \verts, d)}(\Net_{\verts}=\net_{\verts})
\;\propto\; \exp\left(\displaystyle\sum_{i\in\verts < j\in\verts} \eta_{i,j}(\params, \verts, d)\; \edg{i,j}\right),
\eeq
where 
\beq
\eta_{i,j}(\params, \verts, d)
\;=\; \param_1 - \param_2\, f(d(i,j)).
\eeq
Here,
$d: \verts \times \verts \mapsto \mR_0^+$ is a distance function and $f: \mR_0^+ \mapsto \mR_0^+$ is a function of distance,
where $\mR_0^+ = \mR^+ \cup \{0\}$ denotes the set of positive real numbers $\mR^+$ and $0$.
The function $f$ specifies how the distance affects the log odds of the probability of an edge:
\beq
\label{conditional.log.odds}
\log \dfrac{\M_{\verts,\cnmap(\params,\verts, d)}(\Edge{i,j} = 1)}{1-\M_{\verts,\cnmap(\params,\verts, d)}(\Edge{i,j}=1)}
\;=\; \param_1 - \param_2\; f(d(i, j)).
\eeq
The function $f: \mR_0^+ \mapsto \mR_0^+$ can take many forms:
e.g.,
if $f(d(i,j)) = d(i,j)$ and $d(i,j)$ is the Euclidean distance between the positions of $i\in\verts$ and $j\in\verts$ in $\mR^d$,
the model is equivalent to the latent space model of \citet{HpRaHm01} with $\param_2=1$ and observed distances.
But other choices of $f: \mR_0^+ \mapsto \mR_0^+$ are possible,
allowing the log odds to decay slower or faster.
\citet{butts.acton:ch:2011} showed that the rate of decay can have a considerable impact on the structure of the population graph,
so care must be taken when specifying ERGMs with spatial structure.
In addition,
ERGMs can contain more model terms depending on distance,
although the log odds of the probability of an edge \eqref{conditional.log.odds} needs to be replaced by the conditional log odds of the probability of an edge when the model terms induce dependence between edge variables.

ERGMs with spatial structure have at least three advantages.
First,
in many applications with spatial structure edges between population members separated by short distances are more likely than edges between population members separated by large distances.
ERGMs with spatial structure offer many opportunities to test and model the impact of spatial structure on the population graph.
Second,
ERGMs assuming that the probability of an edge decreases as a function of distance can induce sparsity by penalizing edges between pairs of nodes that are separated by large distances.
Last,
but not least,
forcing the dependence of edge variables to decay as a function of distance can help control dependence,
resulting in better-behaved ERGMs.  
For instance, 
\citet{Bu10} shows that even a local triangle term based on triangles within a specified radius can be well-behaved provided that the radius is not too large relative to the population density.  
The same idea can be applied to GWESP terms and other geometrically weighted model terms in curved ERGMs,
which are expected to behave even better than local triangle terms.
 
\subsection{ERGMs with temporal structure}
\label{temporal.ergms}

Many networks change over time,
and temporal structure can help construct well-behaved ERGMs:
modeling a sequence of small changes of a large network may be easier than modeling the whole, large network observed at one time point.
\citet{HaFuXi10} introduced discrete-time Markov models to do so,
with transition probabilities parameterized by ERGMs.
\citet{KrHa10} elaborated a separable parameterization,
separating the edge formation and dissolution process.
We do not further discuss such models here,
because ERGMs with temporal structure are a complex class of models that deserves a separate treatment elsewhere, 
and because many of the associated statistical issues are special cases of classes treated elsewhere in this paper.
We refer to the cited literature for details and \citet{RoPa01} and \citet{ouzienko2011decoupled} for some earlier work on temporal ERGMs.

\subsection{ERGMs with non-random and random attributes}
\label{attribute.ergms}

In addition to edges,
nodes may have attributes,
which may be non-random or random,
observed or unobserved.
ERGMs with attributes do not necessarily impose much structure,
but help capture heterogeneity, 
observed or unobserved, 
in the propensities of nodes to form edges and other subgraph configurations.
As such,
these models play an important role,
even though their primary purpose is not to address the lack of structure of simplistic ERGMs.

We first discuss ERGMs with observed non-random and random attributes and then ERGMs with unobserved, random attributes.

\subsubsection{Observed non-random and random attributes}

The attributes of nodes may either be exogenous, non-random (e.g., race) or endogenous, random (e.g., political preference).
Incorporating non-random attributes as predictors of edges makes sense,
is straightforward, 
and has a long tradition in the ERGM literature.
Some examples are provided by \citet{ergmtermsjss} and \citet{HuGoHa08}:
e.g.,
homophily or similarity with respect to categorical attributes of nodes can be captured by including sufficient statistics of the form $\sum_{i<\verts\, <\, j\in\verts} 1(x_i=x_j)\, \edg{i,j}$,
where $x_i$ and $x_j$ are categorical attributes of nodes $i$ and $j$,
respectively,
and $1(x_i=x_j) = 1$ if $x_i=x_j$ and $1(x_i=x_j) = 0$ otherwise.

A special case of interest is when the attributes are random,
governed by a joint probability model for both the random attributes and the random graph.
Models for both random attributes and random graphs were explored by \citet{FeHa12} in the exponential-family framework.
The resulting models are complex models, and we refer to \citet{FeHa12} for details.

\subsubsection{Unobserved random attributes}

We distinguish ERGMs with discrete and continuous unobserved random attributes.

The discrete case was considered by \citet{Ko09c},
\citet{ScHa13},
and \citet{Waetal18},
as discussed in Section \ref{block.ergms}.
The continuous case was considered by \citet{thiemichen2016bayesian},
who developed a class of ERGMs with random effects.
In the special case of dyad-independent ERGMs (Section \ref{dyadindependence.ergms}),
there has been a long tradition of using random effects models,
dating back to the $p_2$-models of \citet{Du95} and \citet{DuSnZi04} and the related models of \citet{GiSw04},
which are random effects versions of the $p_1$-models of \citet{HpLs81} and related to the random effects and mixed effects models of \citet{Ho03,Ho05,Ho08}.
\citet{thiemichen2016bayesian} considered more general ERGMs with the number of edges $f_i(\net_{\verts}) = \sum_{j \in \verts,\, j\neq i} \edg{i,j}$ of nodes $i \in \verts$ as sufficient statistics,
along with other sufficient statistics.
\citeauthor{thiemichen2016bayesian} assumed that the weights $\eta_i(\params, \verts)$ of the sufficient statistics $f_i(\net_{\verts})$ are random coefficients of the form $\eta_i(\params,\verts) = \param_1 + \epsilon_i$,
where $\param_1 \in \mR$ can be interpreted as the overall propensity to form edges in the population,
$\epsilon_i \in \mR$ can be interpreted as the deviation of node $i \in \verts$ from the overall propensity,
and $\epsilon_i \iid N(0, \sigma^2)$ ($\sigma^2 > 0$).
The resulting random effects ERGMs can capture unobserved heterogeneity in the propensities of nodes to form edges along with other network features.

\section{Complete- and incomplete-data generating process}
\label{generating.processes}

The discussion of the likelihood function in Section \ref{clarification.likelihood} demonstrates that likelihood-based inference requires proper statistical language to clarify the core statistical notions of ``sample" and ``population" in the ERGM framework,
and to separate the process that generates the population graph from the observation process.

We follow here the principled approach of \citet{Ru76} and distinguish the complete-data generating process (generating the population graph) from the incomplete-data generating process (the observation process).
A failure to take both of these processes into account can lead to misleading statistical conclusions,
as discussed by \citet{Ru76}, \citet{DaDi77}, \citet{ThFr00}, \citet{GiHa06}, \citet{HaGi09}, \citet{KoRoPa09}, \citet{GiHa17}, \citet{crane2018probabilistic}, and \citet{CrDe15}.
We discuss complete- and incomplete-data generating processes in Sections \ref{complete} and \ref{incomplete},
respectively.

The specification of the complete-data generating process serves at least two additional purposes.
First,
the parameters of the complete-data generating process constitute the natural target of statistical inference.
Second,
the population graph or super-population of population graphs generated by the complete-data generating process is the population or super-population to which statistical inferences generalize. 

In addition,
the complete-data generating process is coupled with the goal of statistical inference.
We distinguish three broad goals of statistical inference:
finite-,
super-,
and infinite-population inference, 
which we define below.
These notions are inspired by the corresponding notions in classical statistics \citep[e.g.,][]{hartley1975super}.
We adapt them here to the statistical analysis of network data.

\s

{\bf Definition. Finite-population inference} is concerned with a finite population of nodes $\verts$ and a fixed population graph $\net_{\verts}$ defined on $\verts$.
It does not assume that the population graph was generated by a population probability model.
The goal is to estimate functions of the population graph (e.g., edge counts or the presence or absence of specific edges).

\s

{\bf Definition. Super-population inference} is concerned with a finite population of nodes $\verts$ and a population graph $\net_{\verts}$ defined on $\verts$.
In contrast to finite-population inference, 
it assumes that the population graph was generated by generative process with a population probability model.
The goal is to estimate the parameters of the population probability model.

\s

{\bf Definition. Infinite-population inference} is concerned with an infinite population of nodes $\verts$ and a population graph $\net_{\verts'}$ defined on $\verts'\subset\verts$ generated by a graph process with a population probability model.
The goal is to estimate the parameters of the population probability model.

\subsection{Complete-data generating process}
\label{complete}

The complete-data generating process is the process that generates the complete data,
i.e., 
the population graph of interest.

It is possible to make no assumptions about the complete-data generating process,
leading to finite-population inference (Section \ref{finite-pop}).
If the process that generates the population graph is of substantive interest,
one may specify a super-population.
The specification of a super-population may assume that the size(s) of the graph(s) are either fixed or limited to a finite range of possible sizes,
leading to super-population inference on models of graphs of the same size or similar sizes (Section \ref{fixed.number.of.nodes}).
An alternative is to make assumptions about how the model behaves as the size and composition of the set of nodes $\verts$ changes,
leading to infinite-population inference on models of sequences of graphs of increasing size (Section \ref{ergm.growth}).
We discuss these cases in turn.

\subsubsection{Finite graphs: finite-population inference}
\label{finite-pop}

In some applications,
it is neither necessary nor desirable to make assumptions about the complete-data generating process.
For example, 
consider the network of sexual relationships between HIV-infected residents and non-infected residents of New York City (NYC) during a specified period,
where the goal is to estimate the number of sexual contacts between HIV-infected and non-infected residents.
The population of interest $\verts$ consists of the residents of NYC and the population graph $\net_{\verts}$ consists of sexual relationships between residents of NYC.
If the whole population graph $\net_{\verts}$ is observed,
the population graph can be used to answer the question of interest by counting the number of sexual relationships between HIV-infected and non-infected residents.
If it is not possible to observe the whole population graph $\net_{\verts}$ but a sample of sexual relationships is generated (as discussed in Section \ref{incomplete}),
then the sample can be used to construct an estimator of the number of sexual relationships between HIV-infected and non-infected residents.
But,
regardless of whether the whole population graph $\net_{\verts}$ is observed,
answering the question of interest does not require any assumption about the complete-data generating process.
In such situations,
finite-population inference is all that is needed to answer the question of interest.

\s

\paragraph{Target of statistical inference.} 

In finite-population inference,
any function of the population graph $\net_{\verts}$ is a legitimate target of statistical inference:
e.g.,
in the sexual network example described above,
researchers may be interested in estimating the number of sexual relationships between HIV-infected and non-infected residents of NYC.
Here, 
model-based inference may neither be necessary nor desirable and design-based inference is all that is needed \citep[][]{kurant.et.al:sigcomm:2012,gjoka.et.al:netscicom:2014,gjoka.et.al:tr:2015}.

A special case where model-based inference based on ERGMs is useful for finite-population inference was considered by \citet{KrMo17}.
\citet{KrMo17} used ego-centric sampling to estimate population-level network features of interest,
then used the estimated population-level network features as sufficient statistics of an ERGM to simulate graphs that are similar to the estimated population-level network features.
To elaborate, 
define
\beq
\params(\attrs_{\verts}, \net_{\verts})
\;\;\defeq\;\; \argmax_{\params'\in\paramf}\; \exp\en(){\langle\cnmap(\params', \verts),\, \genstats(\attrs_{\verts},\, \net_{\verts})\rangle-\lnormc\en(){\params', \verts}}
\eeq
and note that the maximizer $\params(\attrs_{\verts}, \net_{\verts})$ exists and is unique as long as $\genstats(\attrs_{\verts},\, \net_{\verts})$ falls into the relative interior of the convex hull of the set $\{\genstats(\attrs_{\verts},\, \net_{\verts}): \net_{\verts} \in \netset_{\verts}\}$ \citep[][p.\ 151]{BN78}.
The maximizer $\params(\attrs_{\verts}, \net_{\verts})$ is a function of the attributes of population members $\attrs_{\verts}$ and the population graph $\net_{\verts}$ and is hence a legitimate target of finite-population inference.
We note that the maximizer $\params(\attrs_{\verts}, \net_{\verts})$ is equivalent to the maximum likelihood estimate,
but $\params(\attrs_{\verts}, \net_{\verts})$ is not random,
because neither $\attrs_{\verts}$ nor $\net_{\verts}$ are random.
In fact, 
if the whole population graph $\net_{\verts}$ is observed,
then the maximizer can in principle be computed without error,
though in practice one may have to approximate the maximizer by using Monte Carlo maximum likelihood estimates as described by \citet{KrMo17}.
The function $\params(\attrs_{\verts}, \net_{\verts})$ is of interest,
because it can be used to simulate graphs that are similar to the population graph:
by well-known exponential-family properties \citep[][Theorem 5.5, p.\ 148]{Br86},
the expected sufficient statistic $\genstats(\attrs_{\verts}, \Net_{\verts})$ matches the sufficient statistic $\genstats(\attrs_{\verts}, \net_{\verts})$ of the population $\verts$ under $\params(\attrs_{\verts}, \net_{\verts})$.
Thus,
graphs simulated from the ERGM with parameter $\params(\attrs_{\verts}, \net_{\verts})$ will have sufficient statistics that are similar to the population graph in terms of the sufficient statistic $\genstats(\attrs_{\verts}, \net_{\verts})$.

An example of a situation in which one may wish to simulate similar graphs are data privacy settings involving network data.
In other words,
researchers may wish to share network data with others,
while protecting the privacy of population members.
To do so,
researchers can simulate a graph that is similar---but not identical---to the population graph $\net_{\verts}$ in terms of network features $\genstats(\attrs_{\verts}, \net_{\verts})$ \citep{FiSa10,karwa2017sharing}.  
\citet{goodreau.et.al:jss:2008} used the described procedure to create synthetic school networks based on the National Longitudinal Study of Adolescent Health,
some of which are included in \texttt{R} package \texttt{ergm} \citep{ergm.jss}.
These networks are used for educational purposes, 
such as tutorials and workshops, 
and software testing.
In such applications,
it is useful to have network data with realistic structure,
but it is not essential to have exact replications of the original network data.

Last,
but not least,
an example that combines both motivations is model-based imputation, 
where one seeks to impute the states of unobserved edge variables in a fixed population graph. 
Model-based imputation \citep{GiHa06,HaGi09,KoRoPa09} can be performed by estimating an ERGM from an incomplete observation of the fixed population graph (using the likelihood function \eqref{likelihood.incomplete} described in Section \ref{clarification.likelihood}) and then simulating draws from the estimated model conditional on the observed edges.  
Note that here the model is employed to (a) leverage information from observed edge variables to predict the states of unobserved edge variables and (2) ensures that the imputed population graphs have properties that are compatible with the observed data (as discussed above).
The ERGM used to make the model-based imputations need not be data-generating model of the population graph,
as long as it helps impute the states of unobserved edge variables.
In such settings,
it is natural to assess model performance via prediction of held-out data, 
as proposed by \citet{wang.et.al:sn:2016}.

\subsubsection{Finite graphs: super-population inference}
\label{fixed.number.of.nodes}

While in some applications
it may neither be necessary nor desirable to make assumptions about the complete-data generating process,
in other applications the complete-data generating process is of substantive interest.
For example,
neuroscientists may be interested in the probability law that governs connections between regions of the human brain \citep[e.g.,][]{SiMoLa12,brain16,brain17}.
Here,
interest centers on a population probability model that generates finite graphs of the same size or similar sizes:
e.g.,
in the neuroscience application in Section \ref{application},
the population of interest consists of 56 regions of the human brain,
and the goal of statistical inference is to infer the probability law that governs connections between these 56 regions based on 108 brain networks (i.e., 108 replications).

\s

\paragraph{Target of statistical inference.} 

In super-population inference,
the target of statistical inference is the parameter $\params$ of the population probability model that generated the population graph and governs the super-population consisting of all possible population graphs of the same size or a finite range of possible sizes.
We note that even when the whole population graph is observed, 
uncertainty arises from the fact that the parameter $\params$ is unknown.

\subsubsection{Sequences of graphs: infinite-population inference}
\label{ergm.growth}

In both statistical practice and theory,
it is sometimes convenient to consider sequences of graphs of increasing size.
In many such situations,
there is an explicit or implicit assumption that there exists a graph limit---i.e.,
an infinite graph defined on an infinite population of nodes---to which sequences of graphs converge \citep{Lo12}.
We therefore refer to statistical inference based on sequences of graphs of increasing size as infinite-population inference,
despite the fact that researchers in practice may be more interested in subsequences of graphs of finite sizes rather than the graph limit itself.

In statistical practice,
sequences of graphs of increasing size may be meaningful when,
e.g.,
one observes two or more graphs of different sizes and wishes to formulate a model that is invariant in a well-defined sense.
Consider residents of NYC and Seattle,
where two residents are connected by an edge if the residents meet at least twice a month to work out together.
While NYC has more than 10 times as many residents as Seattle,
it is not credible that the expected number of workout partners of NYC residents is more than 10 times larger than the expected number of workout partners of Seattle residents,
as dense Bernoulli$(\pi)$ random graphs assume.
In such situations,
it is convenient to formulate a model of sequences of graphs of increasing size such that the expected number of edges of each node is invariant to network size and consider the two observed graphs---the large NYC exercise network and the small Seattle exercise network---as two observations taken from a sequence of graphs generated by the model.
ERGMs that respect such desiderata have been developed by \citet{KrHaMo11},
\citet{KrKo14},
and \citet{BuAl15}.
In statistical theory,
it is convenient to embed observed data (e.g., an observed graph) into a sequence of data sets of increasing size (e.g., a sequence of graphs of increasing size),
which is a classic approach in statistical theory:
e.g.,
\citet{Le99} suggested

\begin{quote}
\emph{``...to embed the actual situation in a sequence of situations, the limit of which serves as the desired approximation''} \citep[][p.\ 1]{Le99}.
\end{quote}

Sequences of graphs of increasing size can be constructed in many ways,
e.g.,
graphs can grow by adding nodes or subsets of nodes along with edges.
To cover a wide range of sequences of graphs of increasing size, 
including cumulative and non-cumulative sequences,
let $\mA_1,\; \mA_2,\; \dots$ be a sequence of sets of nodes and $\mA_1,\; \mA_2,\; \dots$ be a sequence of sets of nodes satisfying $\verts_k \subseteq \bigcup_{l=1}^k \mA_l$.
Suppose that the sequence of random graphs $\Net_{\verts_1},\; \Net_{\verts_2},\; \dots$ is generated by a sequence of models of the form $\M_{\verts_1,\cnmap(\params, \verts_1)},\; \M_{\verts_2,\cnmap(\params, \verts_2)}, \dots$,
where the natural parameter $\cnmap(\params, \verts_k)$ may depend on the set of nodes $\verts_k$ and the dimension of parameter $\params$ may grow with the size $|\verts_k|$ of $\verts_k$.
Then the generating processes can be described by a sequence of the form
\[
(\verts_1, \attrs_{\verts_1}, \Net_{\verts_1}, \M_{\verts_1,\cnmap(\params, \verts_1)}),\; (\verts_2, \attrs_{\verts_2}, \Net_{\verts_2}, \M_{\verts_2,\cnmap(\params, \verts_2)}),\; \dots
\]

\s

\paragraph{Target of statistical inference.} 

In infinite-population inference,
the target of statistical inference is the parameter $\params$;
note that $\params$ may not be the natural parameter of the exponential family and that the dimension of $\params$ may be infinite,
because it may depend on the number of nodes,
as the $\beta$-models in Section \ref{dyadindependence.ergms} demonstrates.

\subsection{Incomplete-data generating process}
\label{incomplete}

The incomplete-data generating process is the process that,
conditional on the population graph generated by the complete-data generating process,
determines which subgraphs of the population graph are observed.
In the best-case scenario,
the whole population graph is observed,
but in more common scenarios,
some of the edges in the population graph are unobserved.
The two most common reasons for incomplete data are sampling and missing data.
We discuss selected incomplete-data generating processes, 
with an emphasis on sampling designs (Sections \ref{ego}, \ref{edges}, and \ref{subgraph}) and missing data (Section \ref{missing}).
We conclude with some comments on the fundamental concept of ignorability of incomplete-data generating processes for the purpose of likelihood-based super- and infinite-population inference (Section \ref{ignorability}).

\subsubsection{Sampling nodes: ego-centric sampling and link-tracing}
\label{ego}

If a population of nodes $\verts$ is large,
it may not be possible to observe the whole population graph.
A popular solution is to sample edges by using ego-centric sampling \citep{KrMo17} or link-tracing \citep{ThFr00,GiHa06,HaGi09}.
Both sample a subset of nodes $\verts' \subseteq \verts$ and record edges from nodes in $\verts'$ to nodes in $\verts$.

An ego-centric sampling design generates a sample of nodes along with edges as follows \citep{KrMo17}:\s
\begin{enumerate}
\item Generate a probability sample of nodes,
called egos.
\item For each sampled ego,
record edges to connected nodes,
called alters.
\end{enumerate}
\s
A probability sample of nodes can be generated by any sampling design for sampling from finite populations \citep[e.g.,][]{Th12}.

A number of variations of ego-centric sampling designs are possible.
First,
some ego-centric sampling designs identify alters,
so that it is known whether two egos nominated the same alter.
Second,
other ego-centric sampling designs ask egos to report which pairs of alters have edges \citep{SmMaHo72}.
Third,
an important extension of ego-centric sampling is link-tracing.
Link-tracing exploits the observed edges of sampled nodes to include additional nodes into the sample provided the identities of the egos and alters of sampled nodes are known.
One specific form of $k$-wave link-tracing samples nodes and edges as follows \citep{ThFr00}:\s
\begin{enumerate}
\item Wave $l = 0$: Generate an ego-centric sample.
\item Wave $l = 1, \dots, k$:
\begin{enumerate}
\item Add the nodes who are linked to the population members of wave $l - 1$ to the sample.
\item For each added node,
record edges.
\end{enumerate}
\end{enumerate}
\s
Ego-centric sampling can be considered to be a special case of $k$-wave link-tracing with $k = 0$.  
Additional examples of link-tracing are snowball sampling \citep{Go61} and respondent-driven sampling \citep{He97,SaHe04,GiHa10,Gi11}.
Some link-tracing sampling designs,
such as respondent-driven sampling,
may not generate probability samples,
but approximate probability samples when suitable sampling designs are used \citep{kurant.et.al:jsac:2011,Gi11}.  

\subsubsection{Sampling pairs of nodes: edge sampling}
\label{edges}

While ego-centric sampling and link-tracing sample edges indirectly by first sampling nodes and then recording edges of sampled nodes,
one can sample edges directly.
One example is a sampling design that samples spouses from a frame of married couples,
i.e., 
which samples pairs of nodes connected by an edge (here, marriage).
A theoretical treatment of edge sampling can be found in \citet{CrDe16,CrDe15} and \citet{crane2018probabilistic}.

\subsubsection{Sampling subgraphs}
\label{subgraph}

An alternative approach is based on sampling a subset of nodes $\verts' \subseteq \verts$ and collecting information about the whole subgraph $\net_{\verts'}$ of $\net_{\verts}$ induced by $\verts' \subseteq \verts$.
Sampling subgraphs is distinct from ego-centric sampling and link-tracing,
because subgraph sampling collects information about all edges among nodes in $\verts'$ but does not collect information about edges between nodes in $\verts'$ and nodes in $\verts \setminus \verts'$,
which ego-centric sampling and link-tracing do.
The most widely used form of subgraph sampling is multilevel sampling \citep{SnBo99,multilevelnetwork}.
Consider a population of nodes $\verts$ partitioned into subpopulations $\mA_1, \dots, \mA_K$.
Suppose that a subset of subpopulations $\mS \subseteq \{1, \dots, K\}$ is sampled and that the subgraphs $\net_{\mA_k}$ induced by the sampled subpopulations $\mA_k$ with $k \in \mS$ are observed.
A simple example of a multilevel sample is a sample of school classes from a population of school classes,
generated by any sampling design for sampling from finite populations \citep[e.g.,][]{Th12}.
If all students in the sampled school classes are asked to report edges to other students in the same school class,
the subgraphs induced by the sampled school classes are observed.

\subsubsection{Missing data}
\label{missing}

In addition to design-based missingness due to sampling,
there is out-of-design missingness due to, e.g., nonresponse of respondents in network surveys \citep{GiHa06,HaGi09,KoRoPa09}.
Out-of-design missingness is not under the control of researchers,
but is ignorable for the purpose of likelihood-based super- and infinite-population inference under some conditions,
detailed in Section \ref{ignorability}.

\subsubsection{Ignorable incomplete-data generating processes}
\label{ignorability}

An important concept in likelihood-based super- and infinite-population inference given incomplete data is the notion of ignorability due to \citet{Ru76}.
An incomplete-data generating process is ignorable for the purpose of estimating the parameters of the population probability model provided
\s
\bi
\item[(a)] the probability of not observing the value of an edge variable $\Edge{i,j}$ does not depend on the value of $\Edge{i,j}$;
\item[(b)] the parameters of the complete- and incomplete-data generating process are variation-independent \citep{GiHa06,HaGi09,KoRoPa09}.
\ei
\s
A more formal description of ignorable incomplete-data generating processes can be found in Section \ref{sec:likelihood}.

Examples of ignorable incomplete-data generating processes include ego-centric sampling and link-tracing,
edge sampling,
subgraph sampling,
and data missing at random,
but exclude respondent-driven sampling \citep{LuAi14}.
We refer to \citet{GiHa06}, \citet{HaGi09}, and \citet{KoRoPa09} for likelihood-based inference with ignorable incomplete-data generating processes and \citet{LuAi14} for likelihood-based inference with non-ignorable incomplete-data generating processes.
We discuss likelihood-based inference given incomplete data,
generated by ignorable incomplete-data generating processes,
in Section \ref{sec:likelihood}.

\section{Likelihood-based inference given incomplete data generated by ignorable incomplete-data processes}
\label{sec:likelihood}

We describe likelihood-based inference for well-posed ERGMs,
based on\linebreak
incomplete data generated by ignorable incomplete-data processes.
We focus here on the maximum likelihood approach of \citet{HaGi09},
and note that \citet{KoRoPa09} describe a Bayesian approach.
Both of them are based on the principled approach of \citet{Ru76} to likelihood-based inference in incomplete-data scenarios.

To describe the likelihood-based approach of \citet{HaGi09},
denote by $\Design$ the $|\verts| \times |\verts|$-matrix with elements $\Designij \in \{0, 1\}$,
where $\Designij = 1$ if the value $\edg{i,j}$ of $\Edge{i,j}$ is observed and $\Designij = 0$ otherwise ($i < j$, $i \in \verts$, $j \in \verts$; elements on the main diagonal of $\Design$, and below the main diagonal, are undefined).
The matrix $\Design$ can deal with all forms of incomplete observations of the population graph,
whether data are unobserved due to node sampling, 
edge sampling, 
subgraph sampling, 
missing data, or any combination of the aforementioned incomplete-data generating processes.
Let $\dyadset = \{i < j:\; i \in \verts,\; j \in \verts,\; \Designij = 1\}$ be the set of pairs of nodes with observed data and $\net_{\dyadset} = \{\edg{i,j}:\, i < j:\; i \in \verts,\; j \in \verts\}$ be the observed data.
The incomplete-data generating process is called ignorable for the purpose of likelihood-based inference for the parameter $\params$ of the complete-data generating process provided\s
\bi
\item[(a)] $\M_{\balpha}(\Design = \design \mid \Net_{\verts} = \net_{\verts}) = \M_{\balpha}(\Design = \design \mid \Net_{\dyadset} = \net_{\dyadset})$,
where $\balpha$ is the parameter of the incomplete-data generating process (e.g., the elements of $\balpha$ may be sample inclusion probabilities);
\item[(b)] the parameters $\balpha$ and $\params$ of the complete- and incomplete-data generating process are variation-independent in the sense that the parameter space is a product space.
\ei

\s

In other words,
the incomplete-data generating process is ignorable as long as the probability of being unobserved does not depend on the nature of the unobserved data.
\citet{HaGi09} demonstrated that many sampling designs are ignorable,
including ego-centric and link-tracing sampling designs.

If the incomplete-data generating process is ignorable,
the likelihood function factorizes as follows:
\beq
\begin{array}{lll}
\lik(\balpha,\, \params;\, \net_{\dyadset})
&\propto& \displaystyle\sum_{\net_{\verts}\; \in\; \netset_{\verts}(\net_{\dyadset})}\;\;\; \M_{\balpha}(\Design = \design \mid \Net_{\dyadset} = \net_{\dyadset})\, \M_{\verts,\cnmap(\params,\verts)}(\Net_{\verts} = \net_{\verts})\vspace{.33cm}
\\
&\propto& \underbrace{\vphantom{\sum_{\net_{\verts}\, \in\, \netset_{\verts}(\net_{\dyadset})}}\; \M_{\balpha}(\Design = \design \mid \Net_{\dyadset} = \net_{\dyadset})}\;\; \times\;\; \underbrace{\displaystyle\sum_{\net_{\verts}\; \in\; \netset_{\verts}(\net_{\dyadset})} \M_{\verts,\cnmap(\params,\verts)}(\Net_{\verts} = \net_{\verts})},\vspace{.33cm}
\\
&=& \hspace{1.2cm} \lik(\balpha;\, \net_{\dyadset}) \hspace{1.2cm} \times \hspace{1.75cm} \lik(\params;\, \net_{\dyadset})
\end{array}
\eeq
where $\netset_{\verts}(\net_{\dyadset})$ is the subset of graphs $\net_{\verts} \in \netset_{\verts}$ compatible with the observed data $\net_{\dyadset}$.

As a consequence,
as long as the incomplete-data generating process is ignorable,
likelihood-based inference for\s
\bi
\item the parameter $\balpha$ can be based on the likelihood function $\lik(\balpha;\, \net_{\dyadset})$;\s
\item the parameter $\params$ can be based on the likelihood function $\lik(\params;\, \net_{\dyadset})$.
\ei

\s

As pointed out in Section \ref{clarification.likelihood},
the likelihood function $\lik(\params;\, \net_{\dyadset})$ is based on marginalizations of the population probability mass function,
regardless of whether the model is projective.
Therefore,
the likelihood function $\lik(\params;\, \net_{\dyadset})$ is not affected by lack of projectivity.
In other words,
statistical inference that respects both the complete- and incomplete-data generating process and is based on the likelihood function is not affected by lack of projectivity.
 
Computational methods for likelihood-based inference given incomplete data are described by \citet{HaGi09} and \citet{KoRoPa09}.
Other work on statistical inference given incomplete data can be found in \citet{Sn10}, \citet{PaRoSn13}, \citet{KrMo17}, \citet{karwa2017sharing}, and \citet{GiHa17}.

\section{Consistency and asymptotic normality of estimators}
\label{ergm.consistency}

We review consistency and asymptotic normality results for likelihood-based estimators of well-posed ERGMs in finite-,
super-,
and infinite-population scenarios.
These results demonstrate that likelihood-based inference for well-posed ERGMs is possible.
We do not consider statistical inference for ill-posed ERGMs,
because inferring models which are known to be ill-posed and which are not believed to have generated observed network data is not meaningful.

\subsection{Finite-population inference}
\label{consistency.finite}

Finite-population inference focuses on functions of the population graph,
such as the number of edges in the population graph,
and does not assume that the population graph was generated by a population probability model.
If the whole population graph is observed,
there is no uncertainty.
However,
when a sample from the population graph is generated---as described in Section \ref{incomplete}---there is uncertainty due to the unobserved edges in the population graph. 
In such situations,
two forms of consistency are available for estimators of population quantities based on sample quantities: 
Fisher-consistency and consistency and asymptotic normality under sampling.

First,
many estimators of population quantities are Fisher-consistent \citep{Fi22}.
In other words,
when the whole population graph is observed,
the estimator of the population quantity of interest is equal to the population quantity.
An example is an estimator of the proportion of edges in the population graph based on the proportion of edges in a sample.

Second,
it is often possible to write functions of the population graph of interest in terms of weighted population totals.
In such settings, 
one can construct classical Horvitz-Thompson estimators for the weighted population total of interest, 
whose properties follow from the sampling design and often include consistency and asymptotic normality under sampling \citep{gjoka.et.al:tr:2015}.

Last,
but not least,
consider the following function of the attributes of the population of nodes, $\attrs_{\verts}$, and the population graph, $\net_{\verts}$,
which we motivated in Section \ref{finite-pop}:
\beq
\params(\attrs_{\verts}, \net_{\verts})
\;\;\defeq\;\; \argmax_{\params'\in\paramf}\; \exp\en(){\langle\cnmap(\params', \verts),\, \genstats(\attrs_{\verts},\, \net_{\verts})\rangle-\lnormc\en(){\params', \verts}}.
\eeq
Consider the case where the population quantity $\genstats(\attrs_{\verts},\, \net_{\verts})$ is unknown,
which means that $\params(\attrs_{\verts}, \net_{\verts})$ is unknown.
If an ego-centric sample is generated---as described in Section \ref{ego}---and $\genstats(\attrs_{\verts},\, \net_{\verts})$ is estimated from the ego-centric sample,
then the resulting estimator of $\params(\attrs_{\verts}, \net_{\verts})$ is consistent and asymptotically normal,
provided $\genstats(\attrs_{\verts},\, \net_{\verts})$ can be reconstructed from ego-centric observations of all members of the population $\verts$ \citep{KrMo17}.

\subsection{Super-population inference}
\label{consistency.super}

Super-population inference is concerned with a finite population $\verts$ and a population graph defined on $\verts$,
generated by a population probability model.
We review here concentration and consistency results for likelihood-based estimators of well-posed ERGMs with block structure.
We consider both complete-data scenarios,
where the whole population graph is observed,
and incomplete-data scenarios,
where subgraphs are sampled by ignorable sampling designs.
These concentration and consistency results respect the fact that the population of nodes $\verts$ is finite and cover all finite populations with at least $K \geq 2$ blocks,
although the results are most interesting when $K$ is large.

We assume that the population graph was generated by ERGMs with observed block structure,
as described in Section \ref{block.ergms}.
Examples of populations with observed block structure are populations of students partitioned into school classes,
armed forces partitioned into units of armed forces,
and workforces in companies partitioned into departments.
Suppose that the within-block ERGMs have block-dependent edge and GWESP terms and the between-block ERGMs have block-dependent edge terms.
In other words,
the sufficient statistics of the within-block ERGMs count the number of edges and the number of connected pairs of nodes with $1, \dots, |\mA_k|-2$ shared partners in block $\mA_k$,
and the natural parameters of the within-block ERGMs are
\[
\begin{array}{lll}
\eta_{k,1}(\params, \verts)
\;=\; \param_1\s
\\
\eta_{k,1+m}(\params, \verts)
\;=\; \exp(\vartheta)\, \left[1 - (1 - \exp(-\vartheta))^{m}\right],\;\;
m = 1, \dots, |\mA_k|-2,
\end{array}
\]
where $k = 1, \dots, K$.
Here, 
$\param_2 = \exp(\vartheta) \in (0, 1)$,
so that $\paramf=\mR\times(0,1)$.
We assume henceforth that parameters of the within- and between-block ERGMs are variation-independent, 
i.e.,
the parameter space is a product space.

The following finite-population concentration and consistency results are taken from Corollaries 1 and 2 of \citet{ScSt16}.
The first result assumes that the whole population graph is observed,
whereas the second result assumes that a sample of blocks is generated by an ignorable sampling design and the subgraphs induced by the sampled blocks are observed.
\begin{theorem}
\label{theorem.weaklyprojective1}
Suppose that a finite population of nodes $\verts$ is partitioned into $K$ blocks $\mA_1, \dots, \mA_K$,
where the size of the smallest block is at least $4$ and the size of the largest block is a constant multiple of the smallest block,
and is bounded above by a finite constant.
Let $\params \in \paramf$ be the data-generating parameter and $\widehat\params$ be the maximum likelihood estimator based on a complete observation of the population graph $\Net_{\verts}$.
Then,
for all $\epsilon > 0$, 
there exist $\delta(\epsilon) > 0$ and $C_1 > 0$ such that,
for all $K \geq 2$,
\beq
\M(\norm{\widehat\params - \params}_2\, <\, \epsilon)
\;\;\geq\;\; 1 - 4\, \exp\left(- \delta(\epsilon)^2\, C_1\, K\right),
\eeq
where $\norm{\widehat\params - \params}_2$ denotes the $\ell_2$-distance between $\widehat\params$ and $\params$.
\end{theorem}

More refined, and more general results on maximum likelihood and $M$-estimators,
covering full and non-full, curved exponential-family models of random graph with correct and incorrect model specifications, 
can be found in \citet{ScSt16}.

Theorem \ref{theorem.weaklyprojective1} is a finite-population concentration and consistency result in the sense that it applies to all finite populations with $K \geq 2$ blocks and shows that the probability mass of maximum likelihood estimator $\widehat\params$ concentrates around the data-generating parameter $\params$ provided $K$ is sufficiently large.
Note that these results extend to many other ERGMs with block structure \citep{ScSt16}.
An important special case is given by $K$ independent graphs $\Net_{\mA_1}, \dots, \Net_{\mA_K}$ induced by non-empty, disjoint sets of nodes $\mA_1, \dots, \mA_K$,
where edges between between the sets of nodes $\mA_1, \dots, \mA_K$ are absent with probability $1$.
An example is the $K=108$ human brain networks used in Section \ref{application},
where connections between brains are impossible.
Theorem \ref{theorem.weaklyprojective1} provides a lower bound on the probability of the event $\norm{\widehat\params - \params}_2\, <\, \epsilon$.

Theorem \ref{theorem.weaklyprojective1} assumes that the whole population graph can be observed.
When it is infeasible to observe the whole population graph,
but it is feasible to sample blocks by using an ignorable sampling design and observing the subgraphs induced by the sampled blocks,
then the following finite-sample concentration result can be obtained.

\begin{theorem}
\label{theorem.weaklyprojective3}
Suppose that a sample of blocks $\mL \subseteq \{\mA_1, \dots, \mA_K\}$ is generated by an ignorable sampling design and that the subgraphs of the population graph induced by the sampled blocks are observed.
Let $\params \in \paramf$ be the data-generating parameter and $\widehat\params_{\mL}$ be the maximum likelihood estimator based on the subgraphs induced by $\mL \subseteq \{\mA_1, \dots, \mA_K\}$.
Then,
under the assumptions of Theorem \ref{theorem.weaklyprojective1},
for all $\epsilon > 0$,
there exist $\delta(\epsilon) > 0$ and $C_2 > 0$ such that,
for all $|\mL| \geq 2$,
\beq
\M(\norm{\widehat\params_{\mL} - \params}_2\, <\, \epsilon)
\;\;\geq\;\; 1 - 4\, \exp\left(- \delta(\epsilon)^2\, C_2\, |\mL|\right).
\eeq
\end{theorem}

\vspace{-.75cm}

The difference between Theorems \ref{theorem.weaklyprojective1} and \ref{theorem.weaklyprojective3} is that the total number of subpopulations $|\mK|$ is replaced by the number of sampled blocks $|\mL|$.
If the population is finite but large,
in the sense that the number of blocks $K$ is sufficiently large,
and the number of sampled blocks $|\mL|$ is sufficiently large,
then the probability of event $\norm{\widehat\params_{\mL} - \params}_2\, <\, \epsilon$ is close to $1$.

It is worth comparing these results to \citet{ChDi11}.
\citet{ChDi11} considered infinite populations without additional structure,
resembling Ising models in physics without lattice structure and discrete Markov random fields in spatial statistics without spatial structure,
and allowed edges to depend on many other edges.
Many of the resulting models are ill-posed---as discussed in Section \ref{concern.bad}---and consistent estimation of such ill-posed models may not be possible.
In contrast,
the concentration and consistency results stated above are based on

\s

\begin{itemize}
\item finite populations;
\item populations with additional structure in the form of subpopulations;
\item short-range dependence, in the sense that dependence is restricted to subpopulations;
\item within subpopulations, 
curved exponential-family parameterizations ensure that the added value of additional triangles decreases.
\end{itemize}

\s

Indeed,
the most important implication of Theorems \ref{theorem.weaklyprojective1} and \ref{theorem.weaklyprojective3} is that sensible assumptions give rise to sensible concentration and consistency results for well-posed ERGMs with additional structure in super- and infinite-population scenarios.
The infinite-population case is reviewed in Section \ref{consistency.weak.projectivity}.

\subsection{Infinite-population inference}
\label{consistency.infinite}

We turn to consistency and asymptotic normality results for likelihood-based estimators of well-posed ERGMs in infinite-population scenarios,
including projective ERGMs (Section \ref{consistency.strong.projectivity.ergms}),
dyad-independent ERGMs (Section \ref{consistency.strong.projectivity}), 
and dyad-dependent ERGMs (Section \ref{consistency.weak.projectivity}).
These results cover both projective and non-projective ERGMs,
showing that consistency and asymptotic normality results can be obtained for likelihood-based estimators of well-posed ERGMs despite lack of projectivity.

In addition to the work discussed below,
\citet{XiNe11} showed that consistency results can be obtained under weak dependence assumptions,
but did not give any example of an ERGM with non-trivial dependence that satisfies those weak dependence assumptions;
and \citet{Mu13} established consistency results for the so-called two-star model.

\subsubsection{Projective ERGMs}
\label{consistency.strong.projectivity.ergms}

The first set of consistency results concerns projective ERGMs.

Examples of projective ERGMs are ERGMs with size-invariant natural parameters of fixed dimension of the form
\beq
\cnmap(\params,\verts')
\eqs \params
\;\; \mbox{ for all }\;\; \params\in\{\params\in\mR^p: \psi(\params, \verts') < \infty\}\;\; \mbox{ and all }\;\; \verts' \subseteq \verts,
\eeq
such as Bernoulli$(\pi)$ random graphs with size-invariant edge probability $\pi$ and other dyad-independent ERGMs with size-invariant natural parameters of fixed dimension.
In this setting, \citet[][Theorem 3]{ShRi11} proved the following:
\begin{theorem}
\label{theorem.projective}
Let $\verts_1, \verts_2, \dots$ be a sequence of sets of nodes and $\Net_{\verts_1}, \Net_{\verts_2}, \dots$ be a sequence of random graphs governed by a sequence of projective ERGMs $\M_{\verts_1,\cnmap(\params,\verts_1)}, \M_{\verts_2,\cnmap(\params,\verts_2)}, \dots$,
where $\verts_k = \{1, \dots, k\}$ and $\cnmap(\params,\verts_k') = \params$ for all $\verts_k' \subseteq \verts_k$ ($k = 1, 2, \dots$).
Then the maximum likelihood estimator $\widehat\param_{|\verts|}$ based on $\Net_{\verts}$ is a strongly consistent estimator of $\btheta$ as $|\verts| \to \infty$.
\end{theorem}

The results of \citet{ShRi11} cover dyad-independent ERGMs using attributes of nodes as predictors of edges.

\subsubsection{Dyad-independent ERGMs}
\label{consistency.strong.projectivity}

Most existing consistency and asymptotic normality results concern dyad-independent ERGMs.
Examples are consistency and asymptotic normality results for $\beta$-models and $p_1$-models \citep[][]{DiChSl11,RiPeFi13,KrKo14,YaZhQi15,YaLeZh11,YaWaQi16,Yaetal18,MuMuSe18}.
We present here two interesting examples,
one with node-dependent parameters and one with size-dependent parameters.

The first example concerns $p_1$-models for directed random graphs with node-dependent parameters \citep{YaLeZh11}.
Under $p_1$-models without reciprocity,
the directed edges are independent Bernoulli$(\denspar_{i,j})$ random variables with edge probabilities $\denspar_{i,j} = \ilogit(\alpha_i + \beta_j)$ and natural parameters $\eta_{i,j}(\params, \verts) = \alpha_i + \beta_j$,
where $\params = (\alpha_1, \dots, \alpha_{|\verts|}, \beta_1, \dots, \beta_{|\verts|})$.
To make the model identifiable,
\citet{YaLeZh11} set $\beta_{|\verts|} = 0$,
so that $\params \in \mR^{2\, |\verts| - 1}$.
The following result was shown by \citet[][Theorems 1 and 2]{YaLeZh11}.
\begin{theorem}
\label{p1}
Let $\verts_1, \verts_2, \dots$ be a sequence of sets of nodes and $\Net_{\verts_1}, \Net_{\verts_2}, \dots$ be a sequence of random graphs governed by a sequence of $p_1$-models without reciprocity $\M_{\verts_1,\cnmap(\params,\verts_1)}$, $\M_{\verts_2,\cnmap(\params,\verts_2)}$, $\dots$,
where $\verts_k = \{1, \dots, k\}$ ($k = 1, 2, \dots$).
Assume that $\norm{\params}_\infty \leq \tau \log |\verts|$,
where $0 < \tau < 1\, /\, 44$ and $\norm{\params}_\infty = \max_{1 \leq i \leq 2\, |\verts| - 1} |\param_i|$.
Then
\begin{itemize}
\item with a probability approaching $1$,
the maximum likelihood estimator $\widehat\params_{\verts}$ based on $\Net_{\verts}$ exists, 
is unique, 
and $\norm{\widehat\params_{\verts} - \params}_\infty \ip 0$ as $|\verts| \to \infty$.
\item for any fixed $k \geq 1$,
the vector consisting of the first $k$ elements of $(\widehat\params_{\verts} - \params)$ is asymptotically multivariate normal with mean vector zero  and variance-covariance matrix given by the corresponding $k \times k$ block of the inverse Fisher information matrix as $|\verts| \to \infty$.
\end{itemize}
\end{theorem}

It may be suprising that consistent estimation of the parameter $\params$ of dimension $2\, |\verts| - 1$ is possible. 
Note, however, that the number of independent observations from the $p_1$-model without reciprocity is $|\verts|\, (|\verts| - 1)$,
so the number of independent observations (which is quadratic in $|\verts|$) grows faster than the number of parameters (which is linear in $|\verts|$).
We note that additional results on $p_1$-models with reciprocity and attributes of nodes (e.g., race) as predictors of edges exist \citep[][]{DiChSl11,RiPeFi13,KrKo14,YaZhQi15,YaLeZh11,YaWaQi16,Yaetal18}. 

The second example concerns sparse Bernoulli$(\denspar_{|\verts|})$ random graphs with size-dependent edge probabilities $\denspar_{|\verts|} = \ilogit(\param - \log |\verts|)$ and natural parameters $\eta(\param, \verts) = \param - \log |\verts|$.
The following result is based on Theorem 3.1 of \citet{KrKo14}.
\begin{theorem}
\label{theorem.nonprojective}
Let $\verts_1, \verts_2, \dots$ be a sequence of sets of nodes and $\Net_{\verts_1}, \Net_{\verts_2}, \dots$ be a sequence of random graphs governed by a sequence of sparse Bernoulli random graph models $\M_{\verts_1,\eta(\param,\verts_1)}, \M_{\verts_2,\eta(\param,\verts_2)}, \dots$,
where $\verts_k = \{1, \dots, k\}$ ($k = 1, 2, \dots$).
Then the maximum likelihood estimator $\widehat\param_{|\verts|}$ based on $\Net_{\verts}$ is consistent and $\sqrt{|\verts|}\; (\widehat\param_{|\verts|} - \param)\; \law\; N(0,\, \exp(-\param))$ as $|\verts| \to \infty$.
\end{theorem}

Other consistency and asymptotic normality results for sparse and dense ERGMs with dyad-independence can be found in \citet{KrKo14}.
Sparse ERGMs with dyad-independence are not projective:
e.g.,
sparse Bernoulli$(\denspar_{|\verts|})$ random graphs with size-dependent edge probabilities $\denspar_{|\verts|} = \ilogit(\param - \log |\verts|)$ are not projective,
as shown in Section \ref{concern.nonprojective}.
Therefore,
these consistency and asymptotic normality results demonstrate that,
when meaningful sequences of random graph models are specified and larger graphs contain more information than smaller graphs,
consistency and asymptotic normality results for size-invariant parameters are possible despite lack of projectivity.

\subsubsection{Dyad-dependent ERGMs}
\label{consistency.weak.projectivity}

The following result shows that maximum likelihood estimators of curved ERGMs with block structure are consistent.
The result follows from the super-population concentration and consistency result in Theorem \ref{theorem.weaklyprojective1},
as discussed in Section \ref{consistency.super}.
\begin{theorem}
\label{theorem.weaklyprojective2}
Let $\mA_1, \mA_2, \dots$ be a sequence of blocks,
$\verts_1, \verts_2, \dots$ be a sequence of sets of nodes defined by $\verts_K = \bigcup_{k=1}^K \mA_k$,
$K = 1, 2, \dots$,
and $\Net_{\verts_1}, \Net_{\verts_2}, \dots$ be a sequence of random graphs governed by a sequence of curved ERGMs with block-dependent edge and GWESP terms $\M_{\verts_1,\cnmap(\params,\verts_1)}$, $\M_{\verts_2,\cnmap(\params,\verts_2)}$, $\dots$
Under the assumptions of Theorem \ref{theorem.weaklyprojective1},
the maximum likelihood estimator $\widehat\params_K$ based on $\Net_{\verts_K}$ is a consistent estimator of the data-generating parameter $\params$ as $K \to \infty$.
\end{theorem}

Theorem \ref{theorem.weaklyprojective2} shows that likelihood-based inference for well-posed ERGMs with additional structure and non-trivial dependence structure is possible despite lack of projectivity.
As noted in Section \ref{consistency.super},
these results extend to other well-posed ERGMs with block structure.

\section{Application to human brain networks}
\label{application}

To demonstrate likelihood-based inference for well-posed ERGMs,
we use human brain network data.
A short discussion of how ERGMs can be used in neuroscience applications can be found in the survey paper of \citet{SiBoLa13}.
Some recent applications of ERGMs to human brain network data can be found in \citet{SiHaLa11}, \citet{SiMoLa12}, \citet{brain16}, and \citet{brain17}.
We use here the human brain network data of \citet{brain17}.

\citet{brain17} extracted data from the online PhysioNet BCI data base \citep{physionet, bci},
consisting of EEG recordings from $108$ human subjects.
The EEG recordings cover $56$ regions within each subject's brain,
over four frequency bands in two states,
``eyes closed" and ``eyes open."
We use the same data as \citet{brain17},
focusing on the beta-frequency band in the ``eyes open" state.
The data,
thresholded by \citet{brain17}, 
are binary,
i.e.,
$Y_{i,j}\in\{0, 1\}$,
and undirected,
i.e.,
$Y_{i,j} = Y_{j,i}$ with probability $1$.
Here,
$Y_{i,j} = 1$ can be interpreted as an indicator of a strong connection between brain regions $i$ and $j$.
A more detailed description of the data can be found in \citet{brain17}.
Last,
but not least,
note that the 108 brain networks are fully observed,
although we subsample the 108 brain networks in Section \ref{samplingbrains2} to demonstrate incomplete-data maximum likelihood estimation.

\s

\paragraph{\bf Goal of statistical inference.}
The population of interest consists of 56 regions of the human brain.
The goal of statistical inference is super-population inference:
we want to infer the probability law that governs connections between these 56 regions of the human brain based on 108 replications.  
Note that the size of the human brain is bounded above by the size of the human skull and therefore brain networks cannot grow without bound,
so infinite-population inference based on sequences of graphs of increasing size is not interesting here.

\s

\paragraph{\bf Population probability models.}
As a population probability model,
we use a curved ERGM capturing connectivity and transitivity in the 108 human brain networks.
We compare the curved ERGM  to latent space cluster models,
which capture a stochastic tendency towards transitivity and are the main competitors of curved ERGMs for the purpose of capturing transitivity.
In addition,
we use the Bernoulli$(\pi)$ random graph model with size-invariant edge probability $\pi$ as the primary example of a projective ERGM.
 
\s

\paragraph{\bf Curved ERGM.}
We use a curved ERGM with two levels of multilevel structure,
with level-1 units corresponding to the 56 brain regions and level-2 units corresponding to the 108 brains.
The probability mass function of the curved ERGM is of the form
\beq
\M_{\verts,\cnmap(\params,\verts)}(\Net_{\verts} = \net_{\verts})
= \displaystyle\prod_{k=1}^{108} \M_{\{\verts_k,\verts_k\},\cnmap(\params,\verts)}(\Net_{\verts_k,\verts_k} = \net_{\verts_k,\verts_k}),
\eeq
where
$\verts_k = \{1, \dots, 56\}$ and $\net_{\verts_k,\verts_k}$ is the observed network of brain $k$,
and
\begin{multline*}
\M_{\{\verts_k,\verts_k\},\cnmap(\params,\verts)}(\Net_{\verts_k,\verts_k} = \net_{\verts_k,\verts_k})
\\
\propto \exp\left(\dsum_{m=1}^{9} \eta_m(\params, \verts_k)\, s_m(\net_{\verts_{k}, \verts_{k}}) + \displaystyle\sum_{m=1}^{56-2} \eta_{9+m}(\params, \verts_k)\, \esp_m(\net_{\verts_k, \verts_k})
\right).
\end{multline*}
The sufficient statistics of the model are:\s
\bi
\item $s_1(\net_{\verts_k,\verts_k})$ is the number of edges in brain $k$;\s
\item $s_2(\net_{\verts_k,\verts_k}), \dots, s_8(\net_{\verts_k,\verts_k})$ are the number of nodes with $0, \dots, 6$ edges in brain $k$,
respectively;\s
\item $s_{9}(\net_{\verts_k,\verts_k})$ is the number of paths of length two in brain $k$;\s
\item $\esp_1(\net_{\verts_k, \verts_k}), \dots, \esp_{56-2}(\net_{\verts_k, \verts_k})$ are the number of connected pairs of nodes with $1, \dots, 56-2$ shared partners in brain $k$,
respectively.\s
\ei
\begin{table}[t]
\begin{center}
\begin{tabular}{ll|rr}
\multicolumn{2}{l}{Parameter} & Estimate & Standard error \\ 
\hline  
$\theta_1$    & Edge parameter    & -4.972 & .560  \\
$\theta_9$    & Two-path parameter & -.091 & .033  \\
$\theta_{10}$ & GWESP shift parameter &   .198 & .024  \\
$\theta_{11}$ & GWESP shift parameter &   .305 & .022  \\
$\theta_{12}$ & GWESP base parameter &  1.061 & .018  \\ 
$\theta_{13}$ & GWESP decay parameter &  1.565 & .028 \\
\end{tabular}
\s
\caption{
Monte Carlo maximum likelihood estimates,
including standard errors,
of all parameters in the curved ERGM,
with the exception of the degree parameters $\theta_2, \dots, \theta_8$.
}
\label{tab:curved_estimates}
\end{center}
\end{table}
The natural parameters of the model are:
\beq
\begin{array}{lll}
\eta_{m}(\params, \verts_k)\;=\; \param_m,\;\; m=1, \dots, 9\s
\\
\eta_{10}(\params, \verts_k)
\;=\; \param_{10} + \param_{12}\, \exp(\param_{13})\, \left[1 - (1 - \exp(-\param_{13}))\right]\s
\\
\eta_{11}(\params, \verts_k)
\;=\; \param_{11} + \param_{12}\, \exp(\param_{13})\, \left[1 - (1 - \exp(-\param_{13}))^{2}\right]\s
\\
\eta_{11+(m-2)}(\params, \verts_k)
=\, \param_{12}\, \exp(\param_{13})\, \left[1 - (1 - \exp(-\param_{13}))^{m}\right],\;\;
m = 3, \dots, 56-2,
\end{array}
\eeq
where
\beq
\Theta 
\;=\; \mR^{11}\, \times\, (\mR \setminus \{0\})\, \times \mR.
\eeq
The resulting model is a curved ERGM with a shifted GWESP term,
shifted in the sense that the natural parameters of the numbers of connected pairs of nodes with $1$ and $2$ shared partners are shifted by $\param_{10}$ and $\param_{11}$,
respectively.
If $\param_{10} = 0$ and $\param_{11} = 0$,
the shifted GWESP term reduces to the unshifted GWESP term.
The shifted GWESP term offers more flexibility than the unshifted GWESP term and we found that the shifted GWESP term improves in-sample and out-of-sample performance relative to the unshifted GWESP term.
The model is identifiable as long as $\param_{12} \neq 0$ and the number of nodes is at least $5$, 
so that with positive probability there are connected pairs of nodes with $3$ or more shared partners;
note that $\param_{12}=0$ implies that $\param_{13}$ cannot be estimated,
because $\param_{12}=0$ eliminates the GWESP term.
The number of nodes must be at least $5$,
because $\eta_{10}(\params, \verts_k) = \param_{10} + \param_{12}$ and $\eta_{11}(\params, \verts_k) = \param_{11} + \param_{12}$ when $\param_{13} = 0$,
so adding a constant $c \neq 0$ to $\param_{10}$ and $\param_{11}$ and subtracting $c$ from $\param_{12}$ does not change the likelihood function when the number of nodes is smaller than $5$.
Note that values $\param_{13} < -\log 2$ are identifiable,
but induce a form of model near-degeneracy when $|\verts_k|$ is large,
as explained in Section \ref{clarification.good}.
It is possible to constrain the maximization of the likelihood function to $\param_{13} \geq -\log 2$,
but it is rarely worth enforcing the constraint,
in part because $|\verts_k| = 56$ is small and in part because unconstrained Monte Carlo maximum likelihood algorithms typically do not venture into $(-\infty, -\log 2)$.
A possible explanation is that the probability of network data is much higher on $[-\log 2, +\infty)$ than $(-\infty, -\log 2)$,
where the model is near-degenerate and places low probability mass on graphs that resemble real-world networks,
and consequently the likelihood function will be lower on $(-\infty, -\log 2)$ than $[-\log 2, +\infty)$.

We used {\tt R} package {\tt hergm} \citep{ScLu15} to estimate the curved ERGM by Monte Carlo maximum likelihood methods.
The estimates, 
along with standard errors,
are shown in Table \ref{tab:curved_estimates}.

\s

\begin{figure}[t]
\centering
\begingroup
\setlength{\tabcolsep}{1pt} 
\renewcommand{\arraystretch}{1} 
\begin{tabular}{c}
\includegraphics[scale=.5]{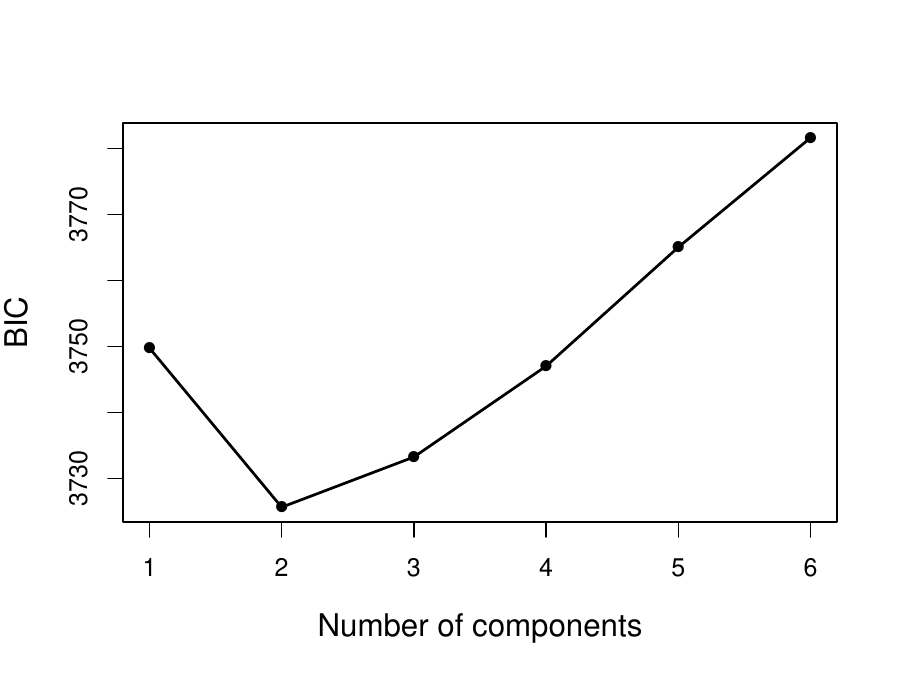}
\includegraphics[scale=.425]{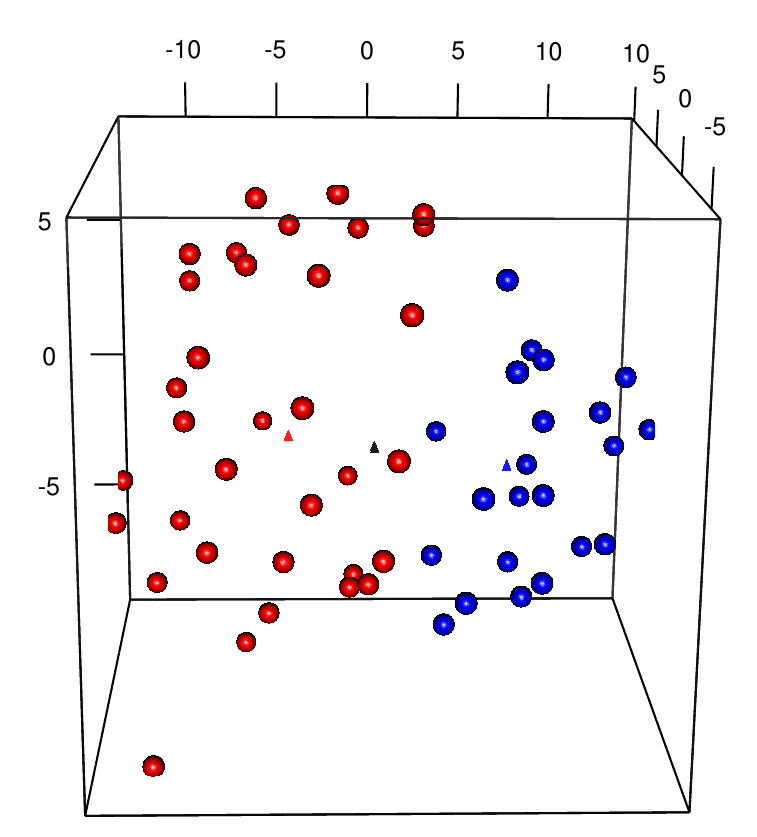}
\end{tabular}
\endgroup
\caption{
Latent space model: 
BIC based on $q=1, \dots, 6$ components (left) and estimated positions of the $56$ nodes in $\mR^3$ based on $q=2$ components (right).
The edges are not shown,
because the three-dimensional plots produced by {\tt R} package {\tt latentnet} can represent binary edges, 
but cannot represent counts of the number of edges between brain regions based on 108 replications.
}
\label{fig:latent_pos}
\end{figure}

\paragraph{\bf Latent space models.}
We use latent space cluster models with node-\linebreak
dependent propensities to form edges,
which generalize $\beta$-models,
stochastic block models,
and latent space models.
Suppose that each node $i$ has a latent position $\bm{z}_i \in \mathbb{R}^3$, 
edges are independent conditional on the positions of nodes,
and the log odds of the conditional probability of an edge between nodes $i$ and $j$ in brain $k$ given the positions of nodes $i$ and $j$ is
\beq
\label{para}
\log
\dfrac{\M_{\verts,\alpha,\bm{\beta}}(Y_{i,j} = 1 \mid \bm{z}_i, \bm{z}_j)}
{1 - \M_{\verts,\alpha,\bm{\beta}}(Y_{i,j} = 1 \mid \bm{z}_i, \bm{z}_j)}
\;=\; \alpha + \beta_{i} \,+\, \beta_{j}  \,-\, \vert\vert \bm{z}_i - \bm{z}_j \vert\vert_{2},
\eeq
where $\alpha \in \mR$ can be interpreted as the overall propensity to form edges and $\beta_i \in \mR$ can be interpreted as the deviation of node $i$.
Since adding a constant $c \neq 0$ to $\alpha$ and subtracting $c/2$ from parameters $\beta_1, \dots, \beta_{56}$ does not change the conditional probability of an edge,
we set $\beta_1=0$.
The positions of nodes are drawn from a multivariate Gaussian mixture distribution with $q$ multivariate Gaussian component distributions,
as in \citet{HaRaTa07}.
The resulting models can be considered as generalizations of the $\beta$-models described in Section \ref{dyadindependence.ergms},
which assume that the log odds of the probability of an edge is $\beta_{i} + \beta_{j}$.
The additional term $-\vert\vert \bm{z}_i - \bm{z}_j \vert\vert_{2}$ can be interpreted as a penalty,
which discourages edges between nodes separated by large distances.
The fact that the positions are generated from a multivariate Gaussian mixture distribution with $q$ components implies that the set of nodes is partitioned into $q$ subsets, 
so one can view the resulting model as a generalization of stochastic block models.
Such latent space models---generalizing $\beta$-models \citep[e.g.,][]{DiChSl11},
stochastic block models \citep[e.g.,][]{NkSt01},
latent space models \citep[e.g.,][]{HpRaHm01},
and latent space cluster models \citep{HaRaTa07}---were proposed by \citet{KrHaRaHo07}.

We used {\tt R} package {\tt latentnet} \citep{latentnet.jss} to estimate the latent space model.
The number of components $q$ was selected by BIC,
as recommended by \citet{HaRaTa07}.
The BIC shown in Figure \ref{fig:latent_pos} suggests to choose $q=2$ components.
All following results are based on $q=2$ components.
The estimated positions of the $56$ nodes in $\mR^3$ based on $q=2$ components can be seen in Figure \ref{fig:latent_pos}.

\s

\paragraph{\bf Bernoulli random graph model.}
We use the Bernoulli$(\pi)$ random graph model with size-invariant edge probability $\pi$ as the primary example of a projective ERGM.
The Bernoulli$(\pi)$ random graph model has probability mass function
\beq
\M_{\verts,\eta(\param,\verts)}(\Net_{\verts} = \net_{\verts})
= \displaystyle\prod_{k=1}^{108} \M_{\{\verts_k,\verts_k\},\eta(\param,\verts)}(\Net_{\verts_k,\verts_k} = \net_{\verts_k,\verts_k}),
\eeq
where
\beq
\M_{\{\verts_k,\verts_k\},\eta(\param,\verts)}(\Net_{\verts_k,\verts_k} = \net_{\verts_k,\verts_k})
\;\propto\; \exp\left(\eta(\param, \verts) \dsum_{i \in \verts_k \,<\, j \in \verts_k} \edg{i,j}\right)
\eeq
with natural parameter $\eta(\param, \verts) = \logit(\pi) = \param \in\mR$.

\s

\paragraph{\bf Other random graph models.}
We do not use $\beta$-models and stochastic block models \citep[e.g.,][]{NkSt01},
because both can be viewed as special cases of the latent space model described above.
In addition,
these and other models are not designed to capture transitivity,
so comparing them to the curved ERGM and latent space model would be unfair to them.

\subsection{In-sample performance}
\label{insample}

\begin{figure}[t]
\centering
\begingroup
\setlength{\tabcolsep}{7pt} 
\renewcommand{\arraystretch}{1} 
\begin{tabular}{ccc}
\hspace{.75cm} 
Bernoulli model & 
\hspace{.75cm} 
Latent space model & 
\hspace{.75cm} 
Curved ERGM\\
\includegraphics[width = .3 \linewidth, height = .2 \textheight, keepaspectratio]{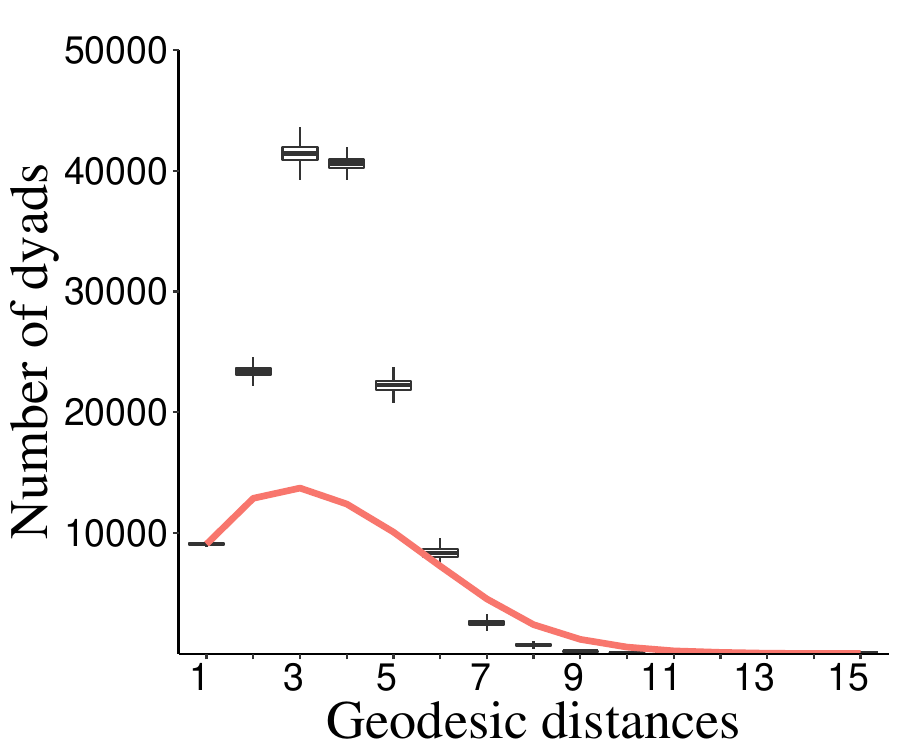}
&
\includegraphics[width = .3 \linewidth, height = .2 \textheight, keepaspectratio]{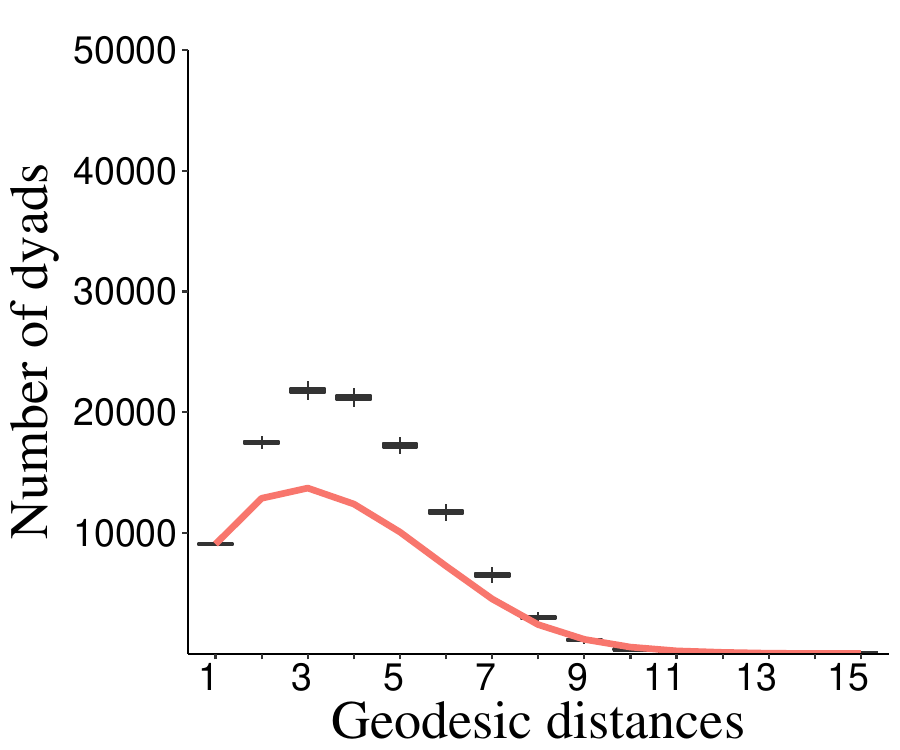}
&
\includegraphics[width = .3 \linewidth, height = .2 \textheight, keepaspectratio]{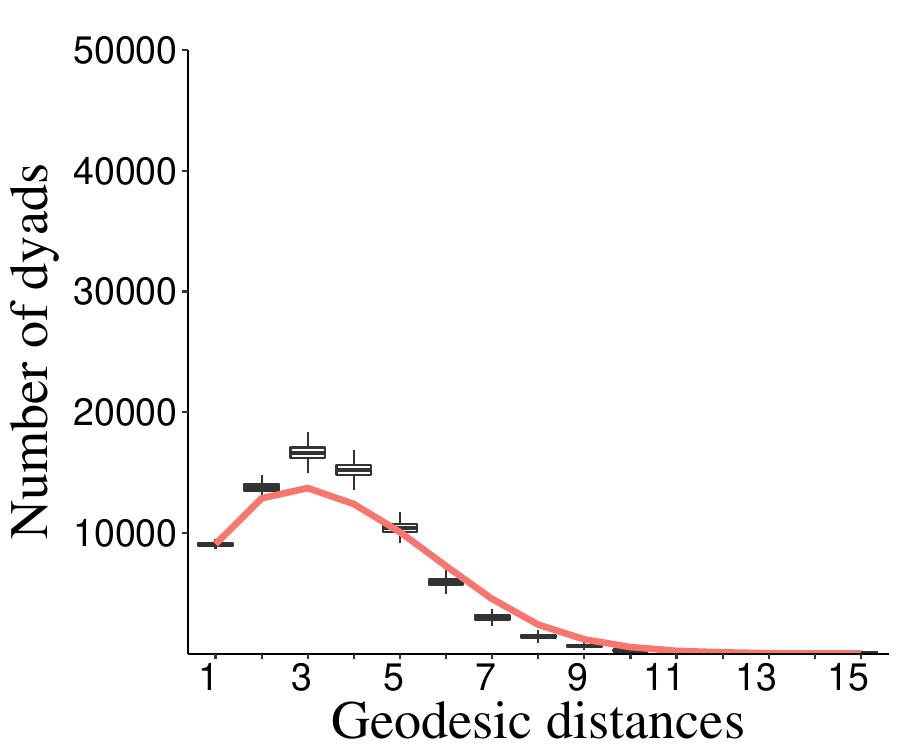}
\end{tabular}
\begin{tabular}{ccc}
\includegraphics[width = .3 \linewidth, height = .2 \textheight, keepaspectratio]{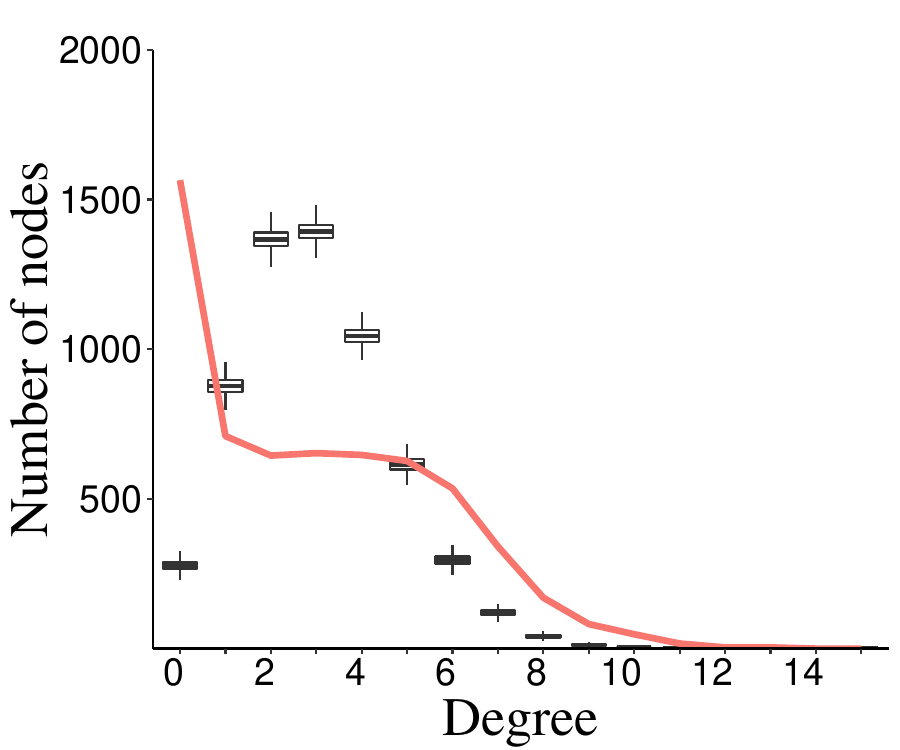}
&
\includegraphics[width = .3 \linewidth, height = .2 \textheight, keepaspectratio]{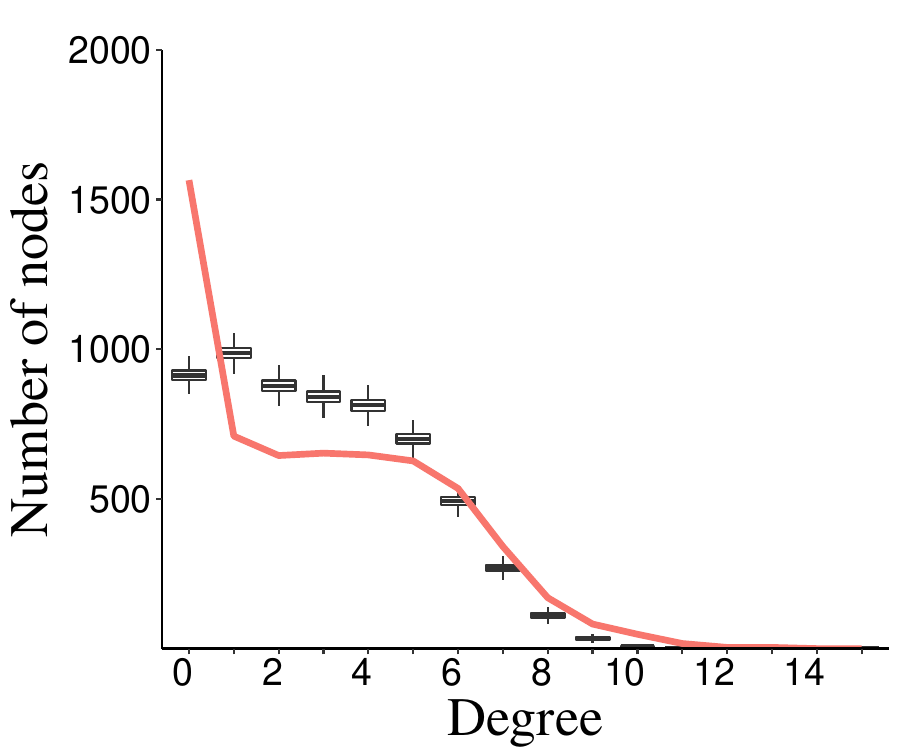}
&
\includegraphics[width = .3 \linewidth, height = .2 \textheight, keepaspectratio]{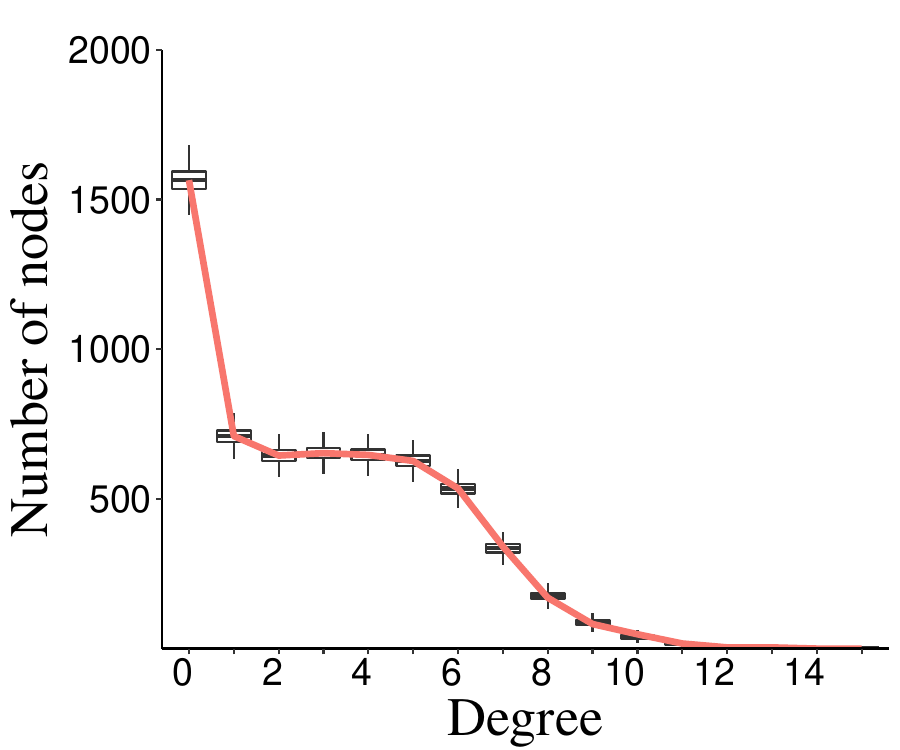}
\end{tabular}
\begin{tabular}{ccc}
\includegraphics[width = .3 \linewidth, height = .2 \textheight, keepaspectratio]{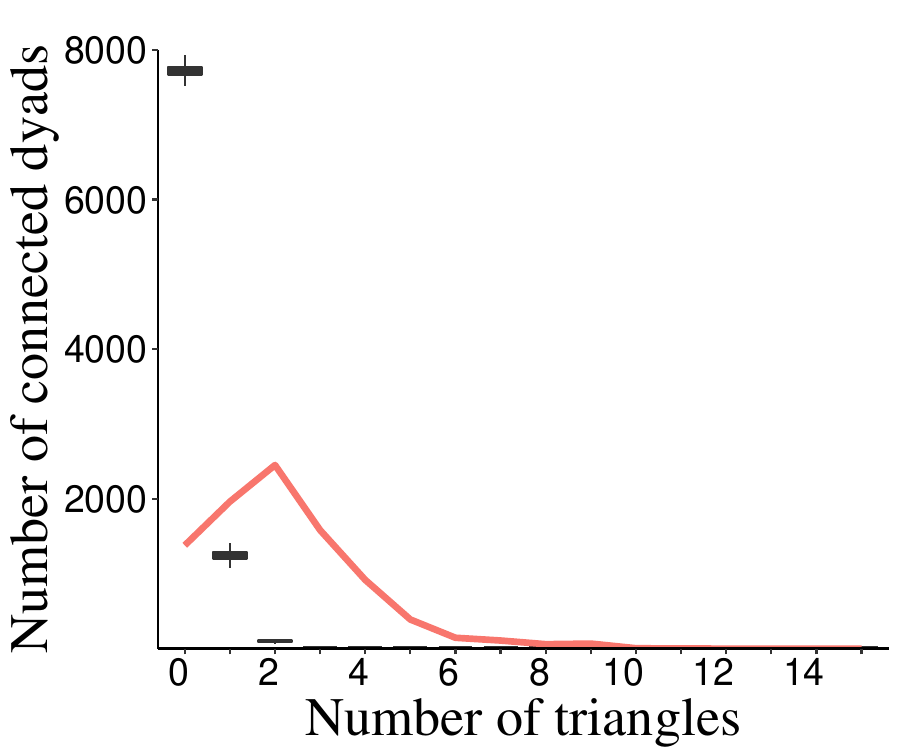}
&
\includegraphics[width = .3 \linewidth, height = .2 \textheight, keepaspectratio]{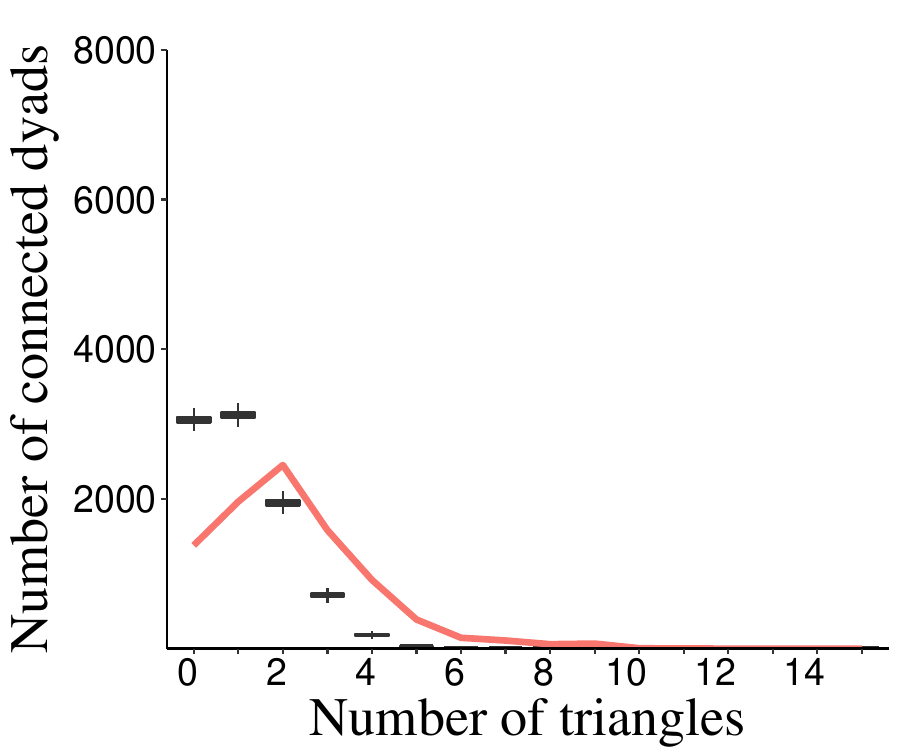}
&
\includegraphics[width = .3 \linewidth, height = .2 \textheight, keepaspectratio]{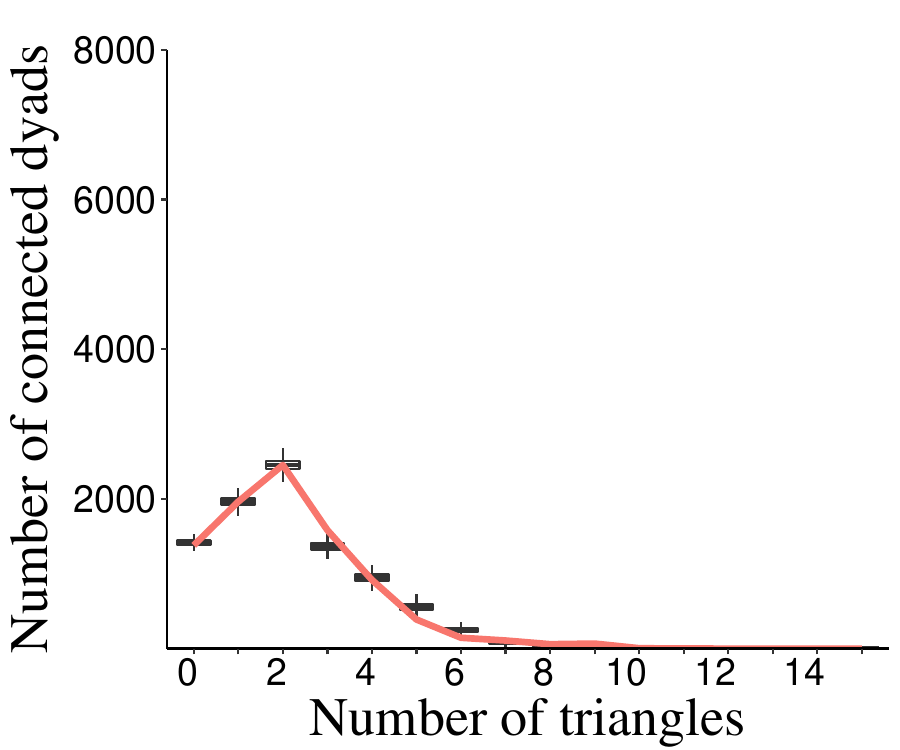}
\end{tabular}
\begin{tabular}{ccc}
\includegraphics[width = .3 \linewidth, height = .2 \textheight, keepaspectratio]{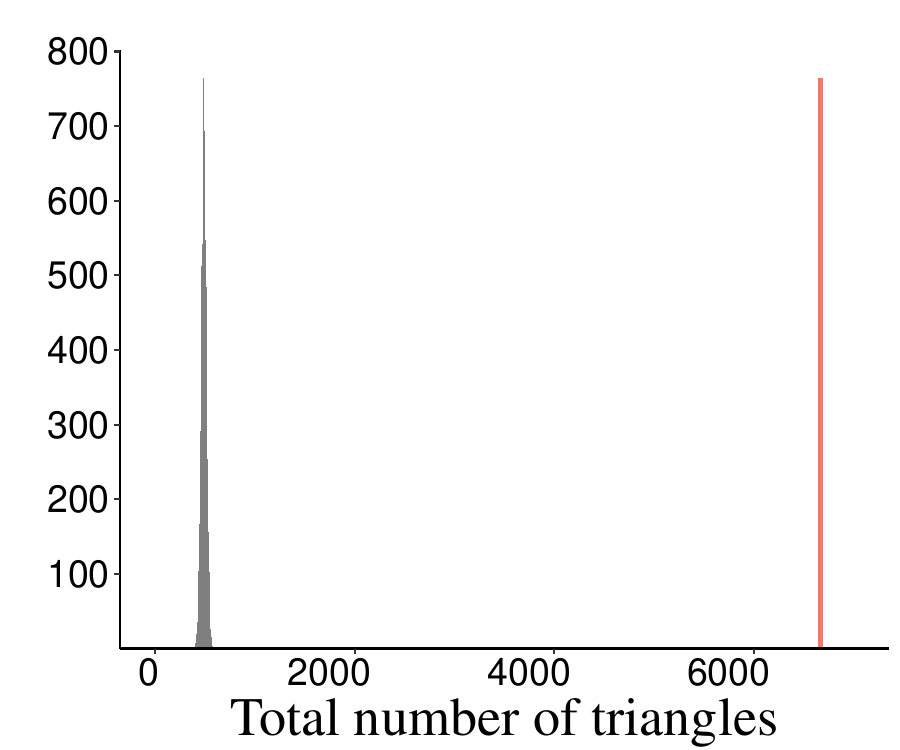}
&
\includegraphics[width = .3 \linewidth, height = .2 \textheight, keepaspectratio]{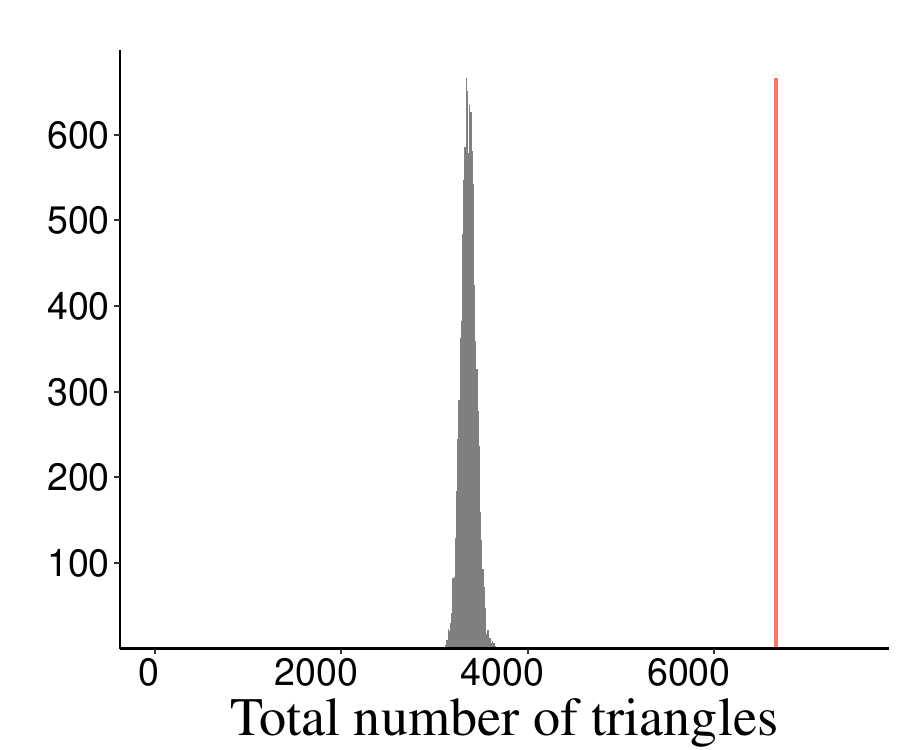}
&
\includegraphics[width = .3 \linewidth, height = .2 \textheight, keepaspectratio]{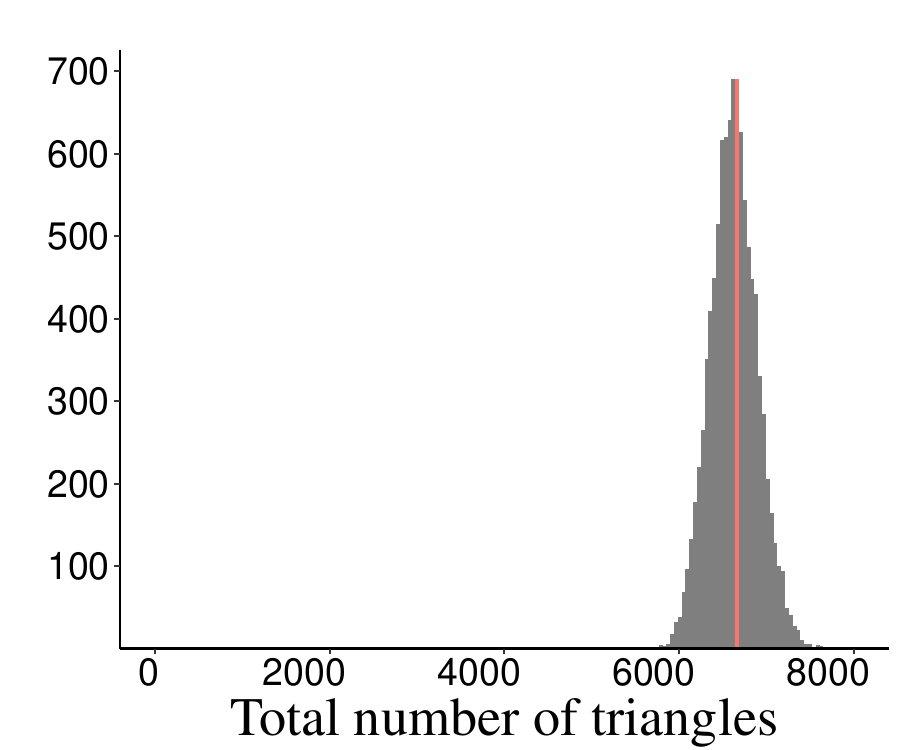}
\end{tabular}
\endgroup
\caption{In-sample performance of the Bernoulli model,
the latent space model,
and the curved ERGM.
The red lines and curves indicate the observed values of the statistics.
}
\label{fig:in_sample}
\end{figure}

We first compare the in-sample performance of the Bernoulli random graph model,
the latent space model,
and the curved ERGM in terms of geodesic distances (the length of the shortest path between dyads);
the number of nodes with degree $m$ (the number of nodes with $m$ edges);
the number of connected dyads with $m$ triangles;
and the total number of triangles;
here,
a dyad refers to a pair of nodes.
All statistics are aggregated over the 108 brain networks.
The first two statistics help assess the in-sample performance of models in terms of connectivity and reachability,
whereas the other two help assess the in-sample performance in terms of transitivity.

The in-sample performance of the three models in terms of these statistics is shown in Figure \ref{fig:in_sample}.
At least two interesting observations can be made.
First,
while some simplistic ERGMs lacking structure are indistinguishable from Bernoulli random graphs in the large-graph limit---as discussed in Section \ref{concern.bad}---the in-sample performance of the curved ERGM is very different from the in-sample performance of the Bernoulli random graph model.
Indeed,
the in-sample performance of the curved ERGM is far superior to the Bernoulli random graph model in terms of both connectivity and transitivity.
Second,
Figure \ref{fig:in_sample} highlights a limitation of latent space models:
while the latent space model induces a stochastic tendency towards transitivity,
it is not a flexible model of transitivity,
and it is not able to match the observed tendency towards transitivity in the observed data.
In fact,
the latent space model has $56$ unrestricted parameters $\alpha$, $\beta_2, \dots, \beta_{56}$ and $56$ latent variables $\bm{z}_1, \dots, \bm{z}_{56}$,
but it is outperformed by the curved ERGM with $13$ unrestricted parameters $\theta_1, \dots, \theta_{13}$,
both in terms of connectivity and transitivity.
Last,
but not least,
it is worth noting that the latent space model is not able to match the observed numbers of nodes with degrees $0, 1, \dots$
While the reasons are unclear,
it is crystal-clear why the curved ERGM is able to match the observed numbers of nodes with degrees $0, \dots, 6$:
the numbers of nodes with degrees $0, \dots, 6$ are sufficient statistics of the curved ERGM and,
under the maximum likelihood estimate,
the expected and observed numbers of nodes with degrees $0, \dots, 6$ are equal.
Note that the numbers of nodes with degrees $7, 8, \dots$ are not sufficient statistics of the curved ERGM,
but the tail seems to be captured by the other model terms.

\subsection{Out-of-sample performance}
\label{outofsample}

\begin{figure}[t]
\centering
\begingroup
\setlength{\tabcolsep}{1pt} 
\renewcommand{\arraystretch}{1} 
\begin{tabular}{cccc}
\includegraphics[scale=.28]{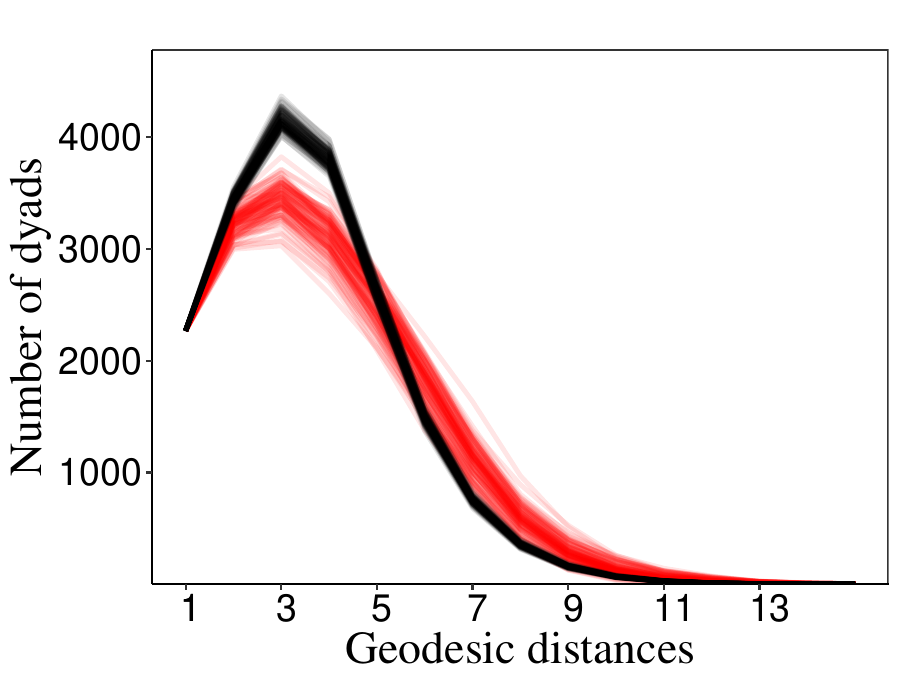}
&
\includegraphics[scale=.28]{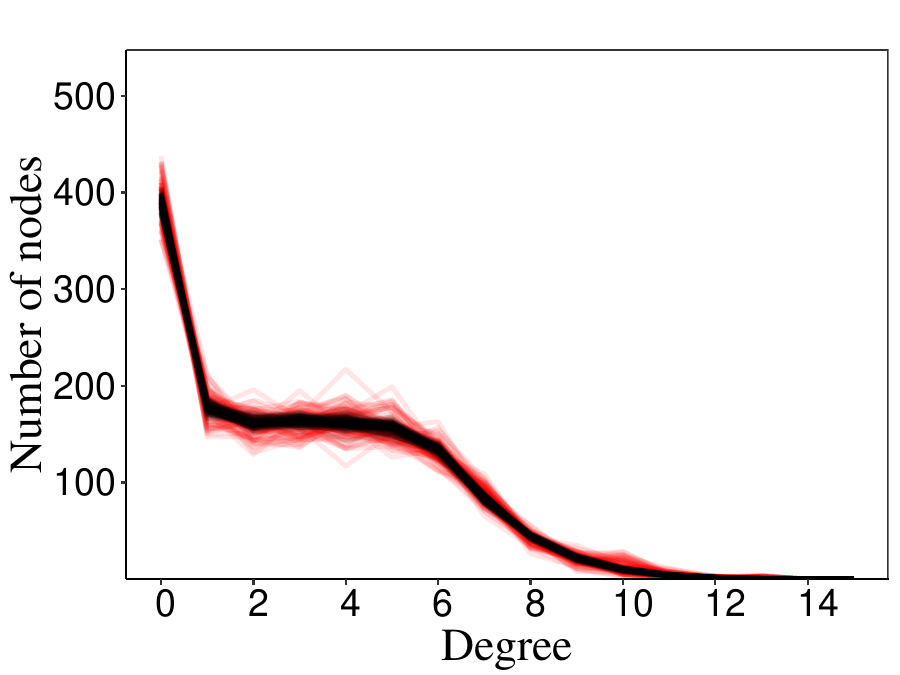}
&
\includegraphics[scale=.28]{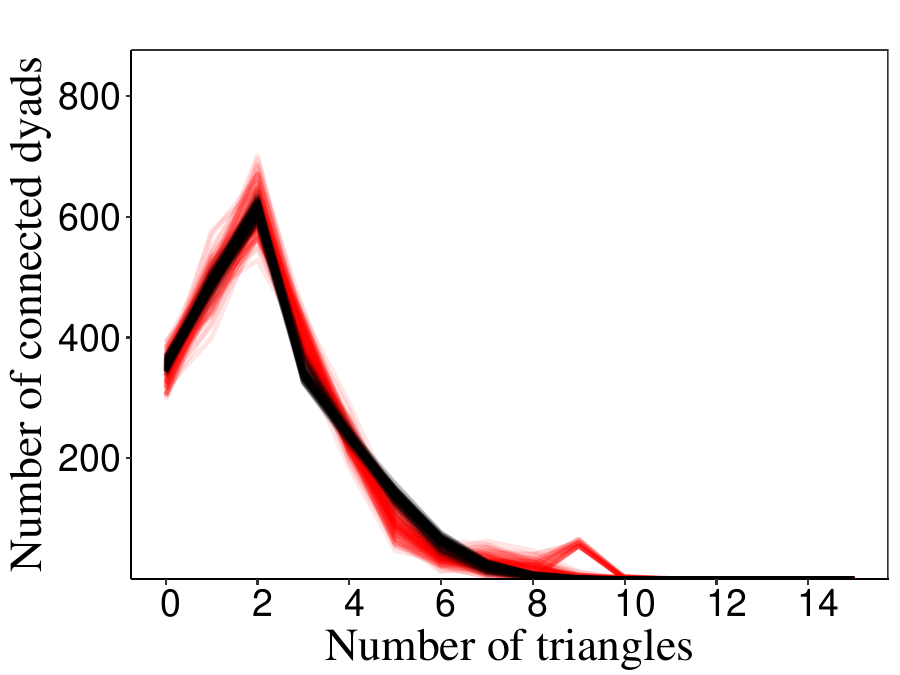}
\end{tabular}
\endgroup
\caption{Out-of-sample performance of the curved ERGM,
using a sample of $75\%$ brain networks to estimate the curved ERGM and $25\%$ of the brain networks to generate model-based predictions.
The black curves are the model-based predictions,
whereas the red curves are the observations.
}
\label{fig:out_of_sample}
\end{figure}

We assess the out-of-sample performance of the best-fitting model,
the curved ERGM,
by sampling $75\%$ of the 108 brains at random and estimating the curved ERGM from the sampled brain networks.
We then generated model-based predictions of the $25\%$ non-sampled brain networks based on the estimated curved ERGM.

Figure \ref{fig:out_of_sample} suggests that the curved ERGM has high predictive power in terms of connectivity and transitivity:
the model-based predictions are close to the observed statistics.
The strong out-of-sample performance lends credence to the assumption that the 108 brain networks were generated by a common data-generating mechanism.

\subsection{Sampling brains: more data helps estimate parameters}
\label{samplingbrains1}

To show that more data helps estimate parameters,
we sampled $25\%$, $50\%$, and $75\%$ of the brains at random and observed the whole brain network of each sampled brain.
We performed the described procedure 50 times.
 
\begin{figure}[t]
\centering
\begingroup
\setlength{\tabcolsep}{1pt} 
\renewcommand{\arraystretch}{1} 
\begin{tabular}{cc}
\includegraphics[width = 0.5 \linewidth, height = .2 \textheight, keepaspectratio]{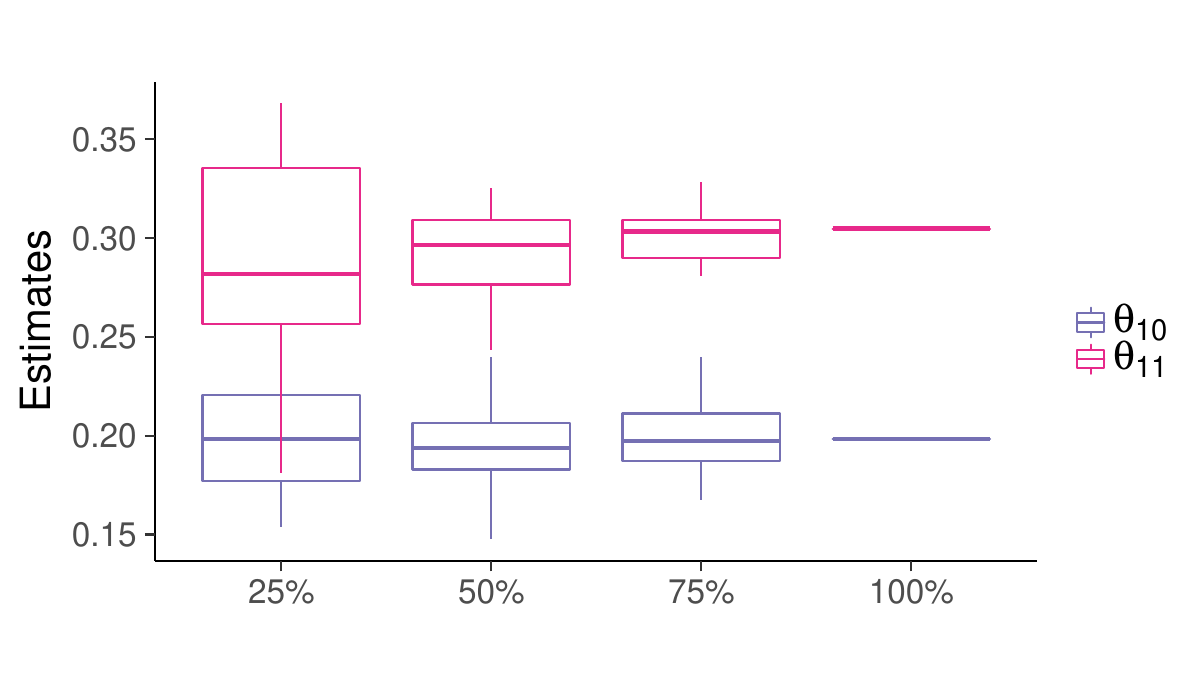}
&
\includegraphics[width = 0.5 \linewidth, height = .2 \textheight, keepaspectratio]{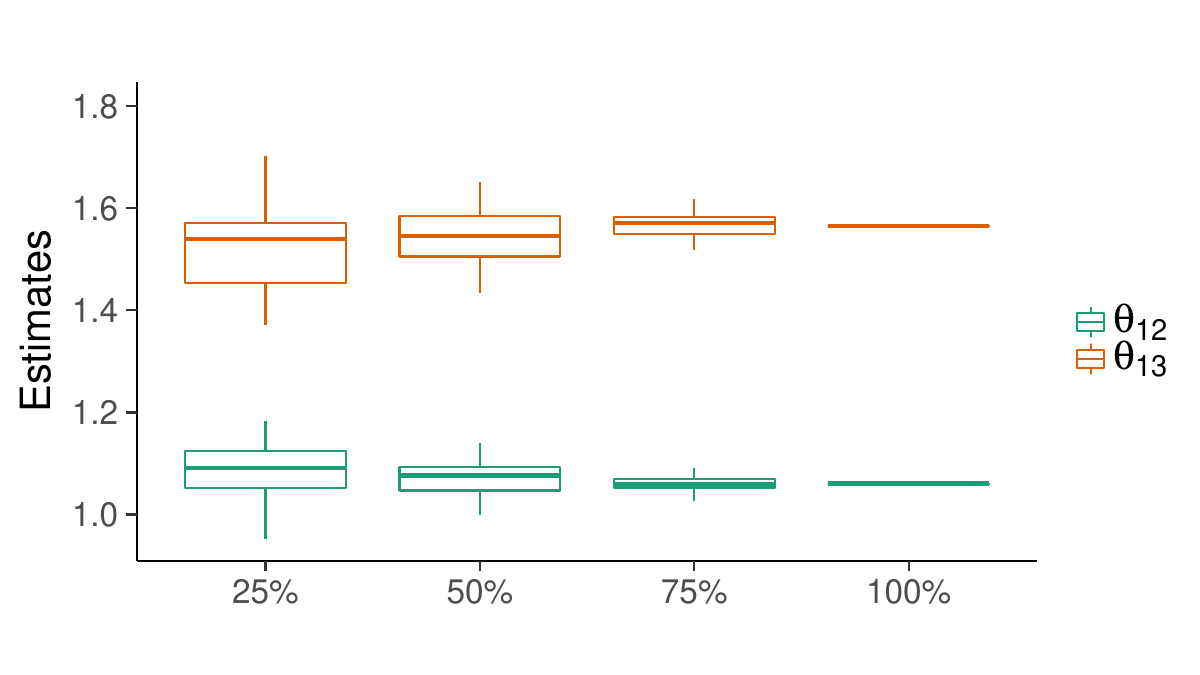}
\end{tabular}
\endgroup
\caption{Sampling brains: Monte Carlo maximum likelihood estimates of the most interesting parameters,
the parameters $\param_{10}, \dots, \param_{13}$ of the shifted GWESP term capturing transitivity,
based on observing $25\%$, $50\%$, $75\%$, and $100\%$  of the 108 brain networks.
}
\label{fig:sub_estimates}
\end{figure}

Figure \ref{fig:sub_estimates} shows that the Monte Carlo maximum likelihood estimates based on samples of brain networks approach the corresponding Monte Carlo maximum likelihood estimates based on all 108 brain networks,
demonstrating that observing more networks does indeed improve our parameter estimates.

\subsection{Subsampling brains: incomplete-data maximum likelihood estimation}
\label{samplingbrains2}

To illustrate likelihood-based inference based on incomplete data generated by ignorable incomplete-data processes,
we sampled 50\% and 75\% of the nodes in each of the 108 brain networks at random and observed the edges of all nodes.
We used the incomplete-data Monte Carlo maximum likelihood procedure described in Section \ref{sec:likelihood} to estimate the parameters from sampled subgraphs.
We performed the procedure 50 times.

\begin{figure}[t]
\centering
\begingroup
\setlength{\tabcolsep}{1pt} 
\renewcommand{\arraystretch}{1} 
\begin{tabular}{cc}
\includegraphics[width = 0.5 \linewidth, height = .2 \textheight, keepaspectratio]{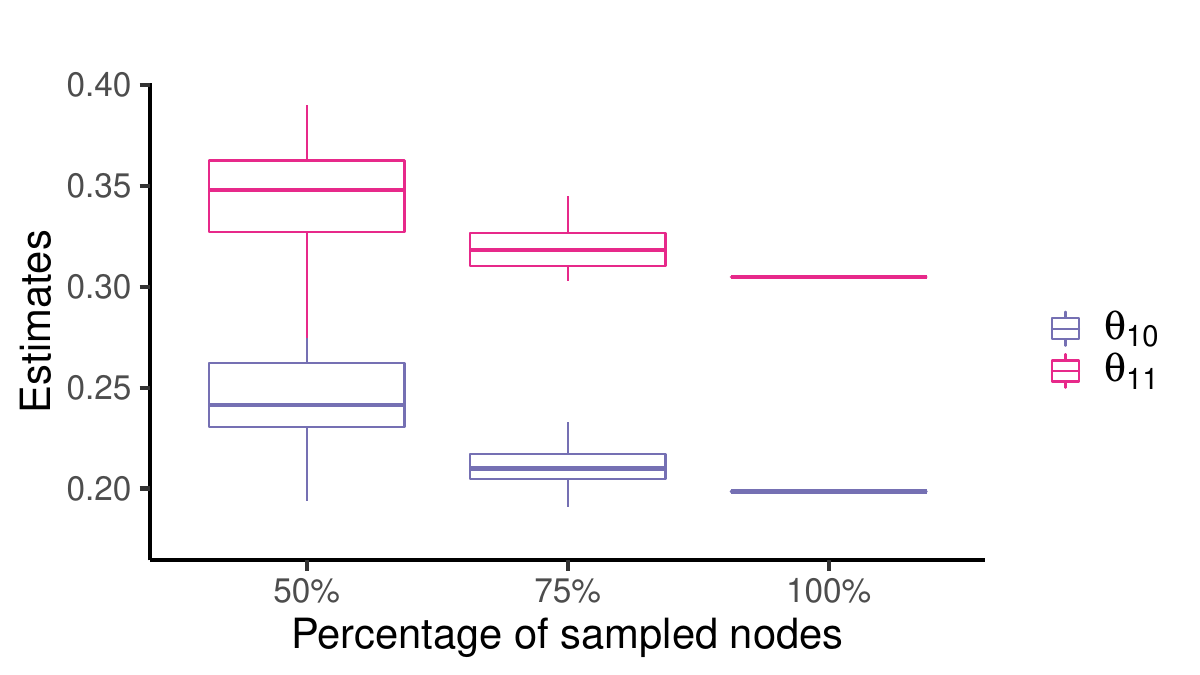}
&
\includegraphics[width = 0.5 \linewidth, height = .2 \textheight, keepaspectratio]{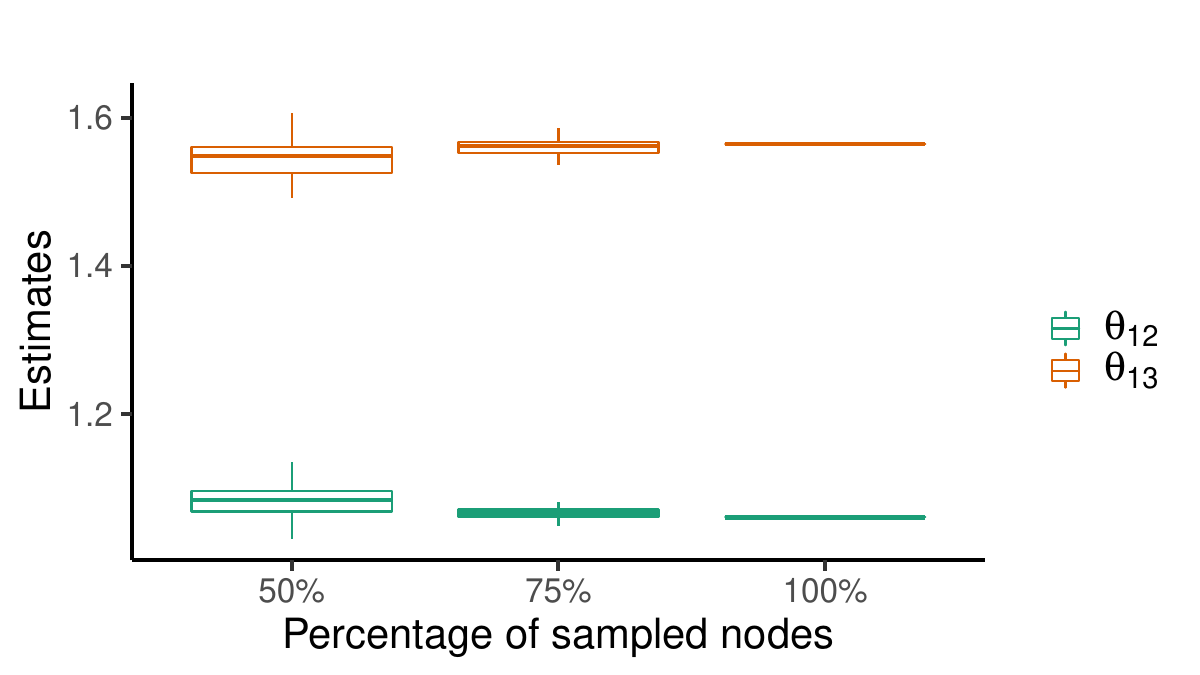}
\end{tabular}
\endgroup
\caption{Subsampling brains: incomplete-data Monte Carlo maximum likelihood estimates of the most interesting parameters,
the parameters $\param_{10}, \dots, \param_{13}$ of the shifted GWESP term capturing transitivity,
based on observing $50\%$, $75\%$, and $100\%$ of the nodes in the 108 brain networks.
}
\label{fig:subsub_estimates}
\end{figure}

The results in Figure \ref{fig:subsub_estimates} show that incomplete-data Monte Carlo maximum likelihood estimates approach the Monte Carlo maximum likelihood estimates based on observing 100\% of the nodes in the 108 brain networks.
These results underscore that statistical inference from subgraphs to population graphs is possible despite lack of projectivity,
as long as statistical inference is based on the likelihood function.

\hide{

Dear Michael,

Thank you for your interest. In attachment you can find two .mat files containing the weighted adjacency matrices of the eyes-open (EO_w) and eyes-closed (EC_w) EEG recordings. The size of each file is 56x56x108X4, the first two dimensions corresponds to the size of the networks (numbers of electrodes), the third one to the number of healthy subjects (108) that we used and the forth one to the different frequency bands (theta, alpha, beta, gamma).

What is measured in entry (i,j) of a specific subject and frequency band, is the spectral coherence between EEG signals of electrodes i and j, in other words measures the statistical dependency over time (one minute recordings) of sensor signals i and j, what is call in the field of brain networks as functional connectivity.

I hope this helps and please do not hesitate to contact us if you need more information or have any questions.

Best regards,
Catalina Obando

}

\subsection{How to deal with graphs of different sizes}
\label{different.sizes}

The human brain network application in Sections \ref{insample}--\ref{samplingbrains2} demonstrates that curved ERGMs can outperform latent space models in super-population scenarios where a population probability model generates graphs of the same size.
A legitimate question to ask is how one can deal with super-population scenarios where a population probability model generates graphs of different sizes.
Such scenarios arise in applications of ERGMs with block structure (Section \ref{block.ergms}) and multilevel structure (Section \ref{multilevel.ergms}).

There are at least three approaches to accounting for different network sizes:\s
\bi
\item[(a)] size-dependent natural parameters using size-dependent offsets \citep{KrHaMo11,KrKo14,BuAl15};
\item[(b)] size-dependent natural parameters using network size as a covariate \citep{slaughter2016multilevel};
\item[(c)] network-specific natural parameters with common mean and network-specific deviations (random effects) \citep{ScHa13}.\s
\ei
All of them assume that the natural parameters have the form 
\begin{quote}\em
natural parameter = size-invariant parameter + size-dependent deviation,
\end{quote}
so that the resulting natural parameters are size-dependent,
and the goal of statistical inference is to estimate the size-invariant parameters.

\section{Conclusions}
\label{conclusion}

The ERGM framework is widely used in practice,
ranging from the study of the human brain and epidemics to differential privacy and social networks (see Section \ref{sec:introduction}).
We believe that the ERGM framework is most useful in super-population scenarios (Section \ref{fixed.number.of.nodes}),
although it can be useful in finite-population scenarios (Section \ref{finite-pop}) and infinite-population scenarios (Section \ref{ergm.growth}),
provided well-posed ERGMs are used and appropriate statistical procedures are employed (e.g., likelihood-based procedures).
The consistency and asymptotic normality results for likelihood-based estimators in finite-,
super-,
and infinite-population scenarios reviewed in Section \ref{ergm.consistency} confirm that statistical inference for ERGMs is possible,
provided the language of exponential families is used to ask well-posed questions.
It goes without saying that the language can be abused to ask ill-posed questions by specifying ill-posed models.
But every language can be abused, 
and potential for abuse does not invalidate its potential for eloquent and effective communication when properly employed.

There is no denying that ERGMs are complex models and give rise to non-trivial computational challenges,
challenges that are shared with other discrete exponential-family models for dependent random variables,
such as discrete Markov random fields in spatial statistics \citep{Bj74,Cr93,St99} and machine learning and statistics \citep[e.g.,][]{RaWaLa10,Yaetal15}.
However,
there is no such thing as a free lunch:
ERGMs model complex dependence,
and modeling complex dependence comes at a price.
Stochastic block models and projective ERGMs are simpler models and more attractive on computational grounds,
but are not capable of capturing the complex dependencies encountered in network data.
Latent space models do capture a stochastic tendency towards transitivity \citep[e.g.,][]{HpRaHm01,HaRaTa07},
but there may be more transitivity in network data than expected under latent space models,
as we found in the human brain network data in Section \ref{application}.
In addition,
latent space models are not flexible models of other forms of complex dependence.
Finally,
likelihood-based inference for latent space models is also expensive in terms of computing time,
even when approximate procedures are used \citep[e.g.,][]{RaNiHoYe12,STMu13}.  
In the end, 
all of these approaches have useful applications and belong to an ever-growing arsenal of statistical tools to understand the structure of complex and dependent network data.

\bibliographystyle{imsart-nameyear}

\bibliography{references}

\end{document}